\documentclass[twocolumn,english,prx,notitlepage,superscriptaddress,longbibliography]{revtex4-2}
\usepackage{graphicx}
\usepackage{amsmath}
\usepackage{amssymb}
\usepackage{bm}
\usepackage{enumerate}
\usepackage{xfrac}
\usepackage[caption=false]{subfig}
\usepackage[normalem]{ulem}
\usepackage[usenames,dvipsnames]{xcolor}
\usepackage{txfonts}
\usepackage[mathscr]{euscript}
\usepackage[pdftex,
pdflang={en-US},
colorlinks=true,
citecolor = blue,
pdfauthor={},
pdftitle={Reconstructing the spatial structure of quantum correlations}]{hyperref}
\usepackage{verbatim}
\usepackage{ulem}

\begin{document}

\title{Reconstructing the spatial structure of quantum correlations in materials}
\author{Allen Scheie}
\email{scheie@lanl.gov}
\affiliation{MPA-Q, Los Alamos National Laboratory, Los Alamos, NM 87545, USA}

\author{Pontus Laurell}
\affiliation{Department of Physics and Astronomy, University of Tennessee, Knoxville, Tennessee 37996, USA}

\author{Elbio Dagotto}
\affiliation{Department of Physics and Astronomy, University of Tennessee, Knoxville, Tennessee 37996, USA}
\affiliation{Materials Science and Technology Division, Oak Ridge National Laboratory, Oak Ridge, Tennessee 37831, USA}

\author{D. Alan Tennant}
\affiliation{Department of Physics and Astronomy, University of Tennessee, Knoxville, Tennessee 37996, USA}

\author{Tommaso Roscilde}
\email{tommaso.roscilde@ens-lyon.fr}
\affiliation{Univ Lyon, ENS de Lyon, CNRS, Laboratoire de Physique, F-69342 Lyon, France}

\date{\today}


\begin{abstract}
Quantum correlations are a fundamental property of quantum many-body states. Yet they remain experimentally elusive, hindering certification of genuine quantum behavior, especially in quantum materials. 
Here we show that the momentum-dependent dynamical susceptibility measured via inelastic neutron scattering 
enables the systematic reconstruction of a general family of quantum correlation functions, which express the degree of quantum coherence in the fluctuations of two spins at arbitrary mutual distance. 
Using neutron scattering data on the compound KCuF$_3$ --- a system of weakly coupled $S=1/2$ Heisenberg chains --- and of numerically exact quantum Monte Carlo data, we show that quantum correlations possess a radically different spatial structure with respect to conventional correlations. Indeed, they exhibit a new emergent length scale --- the quantum coherence length --- which is finite at any finite temperature (including when long-range magnetic order develops). Moreover, we show theoretically that coupled Heisenberg spin chains exhibit a form of quantum monogamy, with a trade-off between quantum correlations along and transverse to the spin chains. 
These results highlight real-space quantum correlators as an informative, model-independent means of probing the underlying quantum state of real quantum materials.
\end{abstract}
\maketitle

\section{Introduction}

Quantum superpositions are among the most profound and fascinating phenomena in nature. 
They lead 
to a variety of quantum correlations, including entanglement \cite{RevModPhys.81.865} and Bell nonlocality \cite{Brunneretal2014}, both considered resources in quantum information processing. Such quantum correlations have been experimentally demonstrated 
in systems isolated from their environment with few degrees of freedom, such as photons \cite{Wangetal2016,Thomasetal2022}, atoms \cite{Monzetal2011,Schmied16,Omranetal2019}, and superconducting circuits \cite{Songetal2019,Mooneyetal2019}. 
However, quantum materials---which host a wealth of exotic physical states \cite{keimer2017physics}---sit at the opposite end of the many-body spectrum. 
They contain Avogadro numbers of quantum-mechanical degrees of freedom, interacting strongly and locally, so that their physics is very sensitive to the underlying system geometry. 
These interacting degrees of freedom produce some very exotic phenomena, which is why quantum materials are so intensely studied \cite{keimer2017physics,tokura2017emergent}. However, despite much attention, the underlying quantum states of quantum materials is often unknown. 
Certifying the quantum superposition nature of such systems, and understanding effects of geometry and dimensionality of interactions on quantum correlations, represent grand challenges for quantum condensed matter physics, as well as new opportunities to understand the role of quantum mechanics in macroscopic systems.

Fortunately, quantum information  theory offers powerful tools for probing quantum superpositions in generic systems in the form of coherence measures \cite{AdessoBC2016,DeChiara2018,Pezzeetal2018,Frerot2023}. 
Here we focus on observable-based measures, which probe coherences of a quantum state when represented on the eigenbasis of an observable, i.e. the non-commutativity between the observable and the density matrix. 
Typically, coherences are studied via interferometric experiments \cite{Pezzeetal2018} and provide the basis of the metrological sensitivity of a quantum state. Unfortunately, interferometry is rarely accessible in the solid-state context; nor is the density matrix itself. 
However, recent works \cite{Hauke2016,FrerotR2016,Frerot2019} have related quantum coherence measures for quantum states in thermal equilibrium to linear response functions, which are directly accessible to spectroscopic techniques  such as light scattering, AC magnetometry, and inelastic neutron scattering. This link allowed neutron scattering experiments on quantum magnets \cite{PhysRevResearch.2.043329, PhysRevB.103.224434,PhysRevLett.127.037201} to reconstruct their quantum Fisher information (QFI) \cite{BraunsteinC1994,Pezzeetal2018}. 
Measurements of QFI associated with order parameters have, in turn, led to estimates of the entanglement depth --- i.e. a lower bound to the minimal number of entangled degrees of freedom in a multipartite entangled state --- in the low-temperature phase of low-dimensional magnets, such as spin chains and triangular antiferromagnets \cite{PhysRevB.103.224434,PhysRevLett.127.037201,Scheie2021}.

\begin{figure}
    \includegraphics[width=\columnwidth]{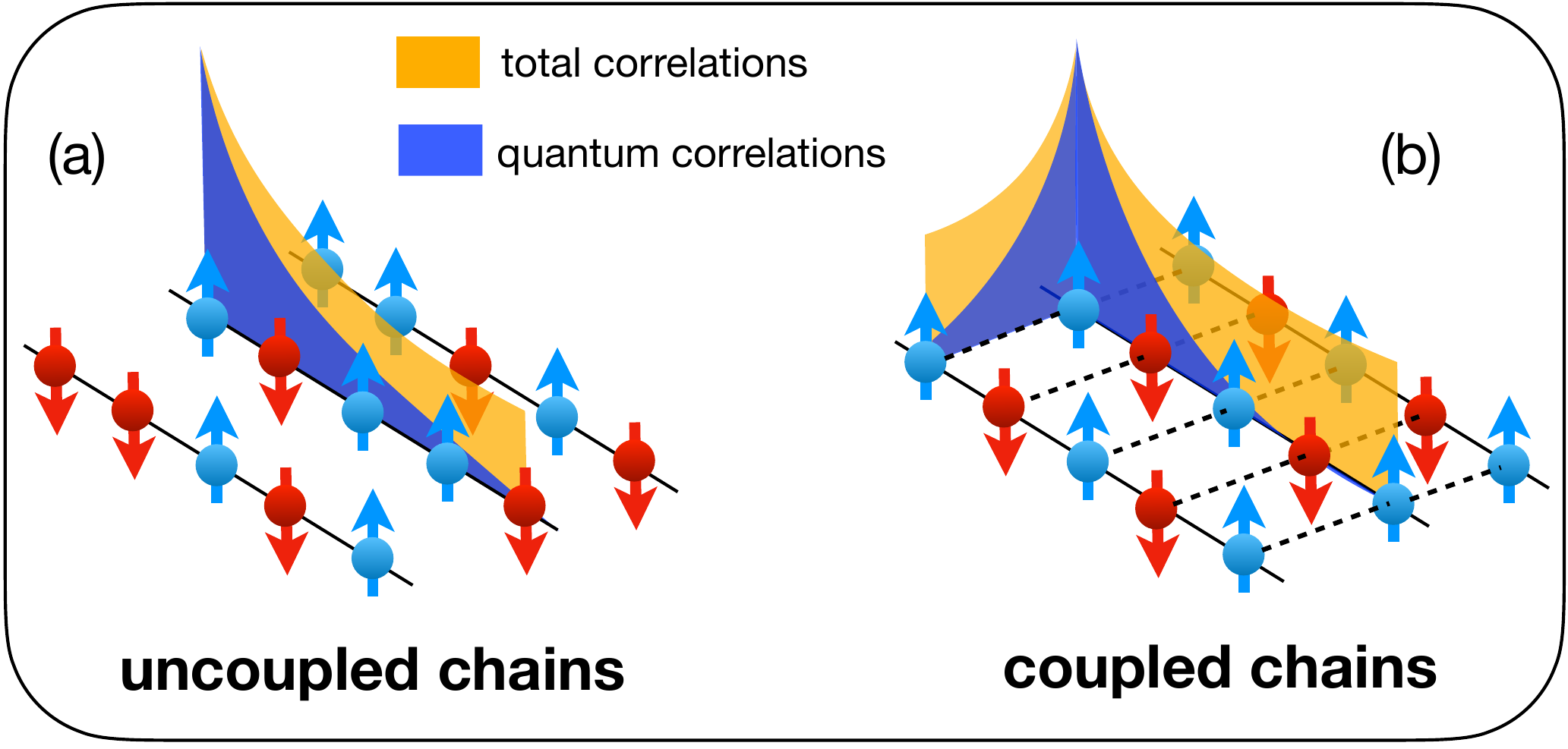}
    \caption{\label{fig:schematic} {Total vs. quantum correlations in $S=1/2$ Heisenberg chains.} (a) Uncoupled 1D chains, and (b) chains subject to interchain coupling. Total and quantum correlations are widely different at any finite temperature. Whereas total correlations are enhanced in all directions when coupling the chains at fixed temperature, quantum correlations are redistributed spatially (at low temperatures), due to an effective form of monogamy (i.e. mutual exclusion). This illustration depicts a two-dimensional system for clarity, but in the paper we consider a three-dimensional system with chains coupled along both directions perpendicular to the chains.}
   \label{f.sketch}
\end{figure}

In this work we show that a Fourier analysis of the linear response function measured in neutron scattering, re-weighted by an appropriate quantum filter function, allows one to extract the full spatial structure of quantum correlations in a model-free manner (i.e., it is applicable to arbitrarily complex systems beyond the reach of simulation techniques). 
Importantly, as we show in this study, such analysis can reveal surprising and new information, even for very well-studied models and materials.


Making use of neutron scattering data on the $S=1/2$ Heisenberg antiferromagnetic chain system KCuF$_3$ \cite{PhysRevB.103.224434, Lake2005, PhysRevLett.111.137205} and quantum Monte Carlo (QMC) simulations, we show that 
these quantum correlation functions share a common spatial structure; and, unlike the ordinary correlation function, they exhibit an exponential decay at all finite temperatures, with an emergent quantum coherence length \cite{MalpettiR2016} which differs substantially from the ordinary correlation length. We provide therefore a clear experimental observation of the short-range nature of quantum correlations at finite temperature, in agreement with recent numerical and analytical results \cite{MalpettiR2016,Frerot2019,PhysRevX.12.021022}. 
We also show numerically that weakly coupled antiferromagnetic chains at low temperature exhibit stronger quantum correlations at short range than strongly coupled chains, due to an effective form of ``monogamy" (i.e. mutual exclusion) of quantum correlations; see Fig.~\ref{f.sketch} for a sketch  summarizing the main results.

\section{Theory of quantum correlation functions} 
Quantum correlation functions can be generally defined as the difference between two types of correlations that are  classically equivalent, and that coincide quantum-mechanically only when the correlated observables commute with the state: \emph{e.g} the statistical correlations of two fluctuating observables, and the response of an observable to a field coupling to the other observable. This latter notion coincides with the \emph{quantum covariance} introduced in Ref.~\cite{MalpettiR2016}; but related quantities (connected to QFI or the Wigner-Yanase-Dyson skew information (SI) \cite{Wigner_1963})  
can also be defined.

For a lattice quantum system, we consider local Hermitian bounded operators $\mathcal{O}_i$, with $i$ the lattice site index; and introduce their sum building up the extensive observable $O = \sum_i O_i$. We then consider the two-site dynamical susceptibility $\chi^{\prime\prime}_{O_i,O_j}(\omega)$ \cite{forster_book},  expressing the out-of-phase variation of the expectation value of $O_i$ in response to a periodic field oscillating at frequency $\omega$ and coupling to $O_j$. Its mathematical expression reads $\chi^{\prime\prime}_{O_i,O_j}(\omega) = -\int dt/\hbar \> e^{-i\omega t} \langle [O_i(t),O_j(0)] \rangle$, where $\langle ... \rangle = {\rm Tr} [(...)\rho]$ represents the thermal equilibrium average at temperature $T$ when $\rho = e^{-\beta H}/{\cal Z}$, with $\beta=(k_B T)^{-1}$, $H$ the system Hamiltonian, and ${\cal Z}$ the partition function. A family of {quantum correlation functions} can then be related to the two-site dynamical susceptibility via an integration over frequency, weighted by an appropriate quantum filter function $h(\beta \hbar \omega)$,
\begin{equation}
C[O_i,O_j;h,\rho] =  \frac{1}{\pi} \int_0^\infty \mathrm{d}\left( \hbar \omega\right)\, h \left( \beta \hbar \omega \right) \chi^{\prime\prime}_{O_i,O_j} \left( \omega \right)~.
\label{eq:C}
\end{equation}
For $C$ to be a well-defined measure of quantum coherence, the function $h(x)$ must satisfy basic mathematical properties \cite{Petz1996,Gibiliscoetal2007,Frerot2017,Frerotetal2022}, namely $h(x) \sim x$ when $x\to 0$, and $h(x \to \infty) = 1$; in this way it acts as a high-pass filter for frequencies $\hbar \omega \gg k_B T$, associated with excitation modes behaving quantum-mechanically at temperature $T$. 
Summing Eq.~\eqref{eq:C} over the spatial indices yields a quantum coherence measure associated with the observable $O$, $I[O;h,\rho] =  \sum_{ij} C[O_i,O_j;h,\rho]$. 
Notable special cases include: 
{
\begin{enumerate}
\item the quantum Fisher information (QFI), $I[O;4h_\mathrm{QFI},\rho] = \mathrm{QFI}(O;\rho)$ for which \cite{Hauke2016,Frerot2017, Frerotetal2022}
 \begin{equation} 
 h_\mathrm{QFI}(x) =\tanh \left( x/2 \right).
 \label{eq:h_QFI}
 \end{equation}
 \begin{align}
     & \mathrm{QFIM}[O_i,O_j;\rho]    = C[O_i,O_j;4h_\mathrm{QFI},\rho]\nonumber \\
      & = \frac{1}{\pi} \int_0^\infty \mathrm{d}\left( \hbar \omega\right)\, 4h_\mathrm{QFI} \left( \beta \hbar \omega \right) \chi^{\prime\prime}_{O_i,O_j} \left( \omega \right)  \label{eq:QFIM}
 \end{align}
 expresses the quantum Fisher information matrix (QFIM) \cite{Liuetal2020}. 
 
\item The quantum variance (Var$_Q$) \cite{FrerotR2016}
$I[O;h_\mathrm{{\rm Var}_Q},\rho] = \mathrm{Var_{Q}}(O;\rho)$ for which
 \begin{equation} 
 h_{\mathrm{Var}_{Q}}(x)=\mathcal{L} \left( x/2 \right),
 \label{eq:h_varQ}
  \end{equation}
where $\mathcal{L}(x)=\coth x - 1/x$ is the Langevin function, and 
\begin{align}
    &\mathrm{Cov}_Q[O_i,O_j;\rho]    =  C[O_i,O_j;h_{\mathrm{Var}_{Q}},\rho]  \nonumber  \\
    & = \frac{1}{\pi} \int_0^\infty \mathrm{d}\left( \hbar \omega\right)\, h_{\mathrm{Var}_{Q}} \left( \beta \hbar \omega \right) \chi^{\prime\prime}_{O_i,O_j} \left( \omega \right)    \label{eq:varQ}
\end{align}
expresses the quantum covariance (Cov$_Q$) \cite{MalpettiR2016,Frerot2019}.

 \item The Wigner-Yanase-Dyson skew information (SI) \cite{Wigner_1963} $I[O;h_\alpha,\rho] = \mathrm{SI}_\alpha(O;\rho)$, for which
 \begin{equation} 
 h_\alpha(x)    =  \frac{ \cosh \left( x/2\right) - \cosh \left[ \left( \alpha -1/2 \right) x \right] } {\sinh \left( x/2\right)},
 \label{eq:h_SI}
\end{equation}
where $0< \alpha < 1$ is a parameter that takes the value of $\alpha=1/2$ in the original Wigner-Yanase definition \cite{Wigner_1963}, with $h_{1/2}=\tanh(x/4)$, and 
\begin{align}
    & \mathrm{SIM}_\alpha[O_i,O_j;\rho] = C[O_i,O_j;h_\alpha,\rho] \nonumber \\
    & = \frac{1}{\pi} \int_0^\infty \mathrm{d}\left( \hbar \omega\right)\, h_\alpha \left( \beta \hbar \omega \right) \chi^{\prime\prime}_{O_i,O_j} \left( \omega \right)   \label{eq:SIM}
\end{align}
expresses the skew-information matrix (SIM). 
\end{enumerate}
}
The Var$_Q$ and ${\rm Cov}_Q$ can  in fact be obtained as the average of $\mathrm{SI}_{\alpha}$ and $\mathrm{SIM}_{\alpha}$ (respectively) on the $\alpha$ index, since $\int_0^1 \mathrm{d}\alpha \, h_\alpha (x) = \mathcal{L}(x/2).$ All the above quantities are intimately linked by the inequality chain (for $\alpha=1/2$) $\mathrm{Var}_Q[O;\rho] \leq \mathrm{SI}_{1/2}[O;\rho] \leq \mathrm{QFI}[O;\rho]/4 \leq  2\mathrm{SI}_{1/2}[O;\rho] \leq 3 \mathrm{Var}_Q[O;\rho]$.


In this work we focus mainly on the quantum covariance, Cov$_Q$, \cite{MalpettiR2016,Frerot2019}. 
This definition corresponds to the difference between static correlations and static response functions \cite{FrerotR2016,MalpettiR2016}, and consequently quantum variance and covariance can be calculated efficiently with QMC. (The QFI and QFIM meanwhile require instead the full reconstruction of the dynamical susceptibility, which is not accessible to QMC directly, due to a notoriously ill-defined analytical continuation of time-dependent correlations from imaginary to real time \cite{JARRELL1996}.)

\begin{figure*}
    \includegraphics[width=\textwidth]{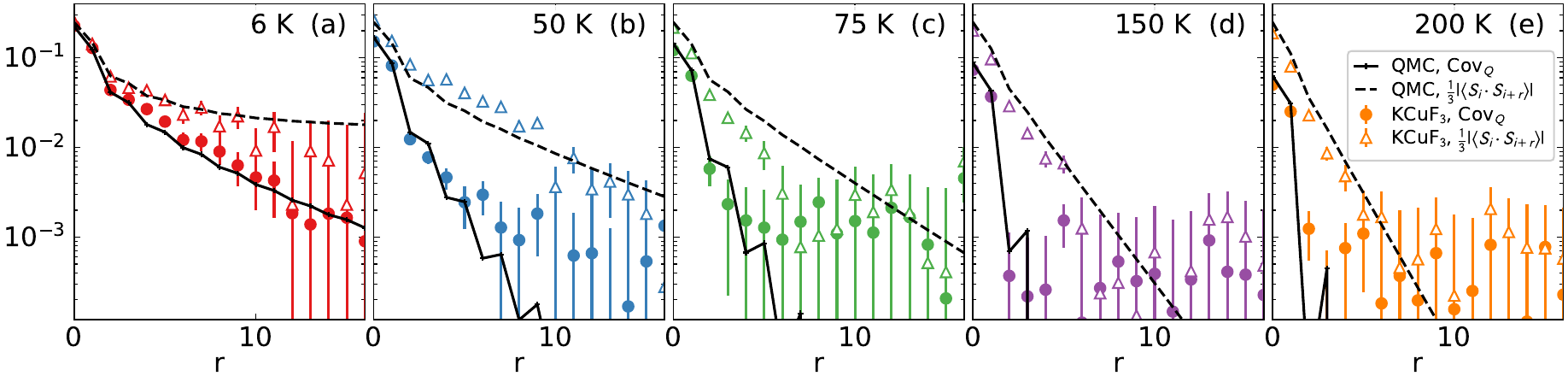}
    \caption{\label{fig:VarQ_vs_T} {Total vs. quantum correlations in KCuF$_3$.}  Reconstructed total and quantum correlations (expressed by the quantum covariance Cov$_Q$ {
    [Eq. \eqref{eq:varQ}]}) 
    along the spin chains of KCuF$_3$ at various temperatures, compared with numerically exact QMC data. The error bars represent one standard deviation uncertainty.}
\end{figure*}

\section{Quantum correlation functions from neutron scattering} 
Inelastic neutron scattering measures the dynamical structure factor $S({\bf Q},\omega)$, related to the momentum-dependent dynamical susceptibility via the fluctuation-dissipation theorem 
{
$
\chi_{\mu\nu}^{\prime\prime}({\bf Q},\omega) = \pi (1 -  e^{- \hbar \omega  \beta})  S_{\mu\nu}({\bf Q},\omega)
$ 
\cite{Lovesey1984}, where  $\chi^{\prime\prime}_{\mu\nu}({\bf Q},\omega) = -\int dt/\hbar \> e^{-i\omega t} \langle {
[ S^\mu_{\bf Q}(t),S^\nu_{-\bf Q}(0) ]} \rangle$; $\mu,\nu = x,y,z$;  and $S^\mu_{\bf Q} = N^{-1/2} \sum_i e^{i{\bf Q} \cdot {\bf r}_i} S_i^\mu$ is the Fourier transform of the $S_i^\mu$ operators for a lattice with $N$ sites.} If $S({\bf Q},\omega)$ is measured across the full Brillouin zone, its inverse Fourier transform allows one to reconstruct the two-site dynamical susceptibility, and calculate the quantum correlation functions. 

To test this idea, we use the neutron scattering data reported in Ref. \cite{Scheie2022} for KCuF$_3$; see Appendix \ref{app:ExperimentalData} for details. This material is an ideal approximation to a system of coupled Heisenberg $S=1/2$ chains \cite{PhysRevLett.111.137205,PhysRevLett.70.4003}
\begin{equation}
H = J \sum_{\langle ij \rangle: \mathrm{chains}} \bm S_i \cdot \bm S_j + J_{\perp}  \sum_{\langle lm \rangle: \mathrm{inter}} \bm S_l \cdot \bm S_m
\label{e.H}
\end{equation} 
where ${\bm S}_i$ is a $S=1/2$ spin operator at site $i$; the first sum runs on nearest-neighbor bonds along the chains; and the second on bonds connecting nearest-neighboring chains to form a tetragonal lattice. 
KCuF$_3$ has weak inter-chain coupling $J_{\perp} = -1.6$~meV compared to the in-chain one $J = 34$~meV \cite{PhysRevB.71.134412}, causing weak 
long-range N\'eel order to appear at a low critical temperature $T_N = 39$~K $\approx 0.1J$. Nevertheless, many salient features of the one-dimensional physics (such as fractional excitations at sufficiently high energy \cite{PhysRevLett.111.137205}) 
are preserved to low $T$ in spite of the long-range ordering. 

In magnetic scattering such as from KCuF$_3$ 
the measured dynamical susceptibility is 
{ $\tilde \chi^{\prime\prime}({\bf Q},\omega) = \sum_{\mu,\nu = x,y,z} (\delta_{\mu\nu} - {\hat{Q}}_{\mu}{\hat{Q}}_{\nu}) \chi^{\prime\prime}_{\mu\nu}({\bf Q},\omega)$} \cite{Squires} 
where ${\hat{Q}}_{\mu}$ are the $\mu = x,y,z$ components of the normalized scattering vector.  In the Heisenberg (isotropic) limit, the spin components act identically and { $\tilde\chi^{\prime\prime}({\bf Q},\omega) = \frac{2}{3} \sum_{\mu = x,y,z} \chi^{\prime\prime}_{\mu\mu}({\bf Q},\omega)$}. 
The two-site spin susceptibility along $\hat{\mu}$, $\chi^{\prime\prime}_{S^\mu_i S^\mu_j} (\omega)$, is obtained by the spatial Fourier transform of $\chi^{\prime\prime}_{\mu\mu}({\bf Q},\omega)$. We also define the two-site dynamical susceptibility $\chi^{\prime\prime}_{ij}(\omega) = \frac{1}{3} \sum_{\mu = x,y,z} \chi^{\prime\prime}_{S^\mu_i S^\mu_j} (\omega)$, including a prefactor $1/3$ for convenience.
{
Thus,
}
the quantum correlation functions defined above can then be reconstructed, along with the total correlation function
\begin{equation}
C_{\rm tot}(i,j) = \frac{1}{3} \langle {\bm S}_i \cdot {\bm S}_j \rangle = \frac{1}{\pi} \int d(\hbar \omega) ~\coth(\beta \hbar\omega/2) ~\chi^{\prime\prime}_{ij}(\omega)~,
\end{equation}
{
where the $\coth$ factor} {
converts dynamic susceptibility back to {
$S({\bf Q}, \omega)$} via the fluctuation dissipation theorem \cite{Lovesey1984}. 
The total correlations are} expected to exhibit an exponentially decaying behavior for $T>T_N$, $C_{\rm tot}(i,j)  \sim \exp(-|i-j|/\xi)$ with $\xi$ the correlation length; while the divergence of $\xi$ at $T_N$ entails the appearance of long-ranged correlations. On the other hand, the quantum covariance ${\rm Cov}_Q$ is expected to exhibit an exponential decay at \emph{any} finite temperature, ${\rm Cov}_Q(i,j)  \sim \exp(-|i-j|/\xi_Q)$ with $\xi_Q$ defining the quantum coherence length, which is finite at any finite temperature, and coincides with $\xi$ only for $T\to 0$. This behavior for Cov$_Q$ has been numerically observed via QMC in Refs.~\cite{MalpettiR2016,Frerot2019}, and only recently it has been {established} as a rigorous result \cite{PhysRevX.12.021022}. Yet an experimental measurement of $\xi_Q$ is still lacking to date.

\section{Quantum correlations for \texorpdfstring{$\rm KCuF_3$}{KCuF3}}
Fig.~\ref{fig:VarQ_vs_T} compares $C_{\rm tot}(i,j)$ and Cov$_Q$ for sites $i, j$ belonging to the same chain, as reconstructed from the neutron scattering structure factor of KCuF$_3$ at various temperatures ($T=$ 6, 50, 75, 150 and 200 K) above and below 
$T_N$. The experimental data are compared with QMC data (obtained via the stochastic series expansion method \cite{SyljuasenS2002}) for a $10 \times 10$ array of 100-site spin chains. 
The experimental results beautifully match the theoretical ones, including the detailed structure clearly visible at short ranges. The vast difference between the {total} correlation length and the emergent quantum coherence length is apparent: while the total correlations go from exponentially decaying (above $T_N$) to decaying to a finite value (below $T_N$), the quantum covariance clearly remains short-ranged at all temperatures, with a decay length $\xi_Q$ significantly smaller than $\xi$. This implies that quantum correlations do not participate in the N\'eel transition (which is not surprising because of the classical nature of the finite-temperature phase transition). 

The asymptotic exponential decay is clearly exhibited by $C_{\rm tot}(i,j)$ for $T>T_N$ and distances exceeding a few lattice steps. 
The spatial structure of the quantum covariance is generally more complex (see Appendix \ref{app:Interchain} 
for an extended discussion); yet a first exponential decay sets in after a few lattice steps, and this decay is clearly visible in the experimental data. We shall focus on the length associated with this short-range decay in the following, and extract it via a linear-regression (LR) estimator  $\xi_{Q,\rm LR}$ from a linear fit of the logarithm of the correlation function 
or via a second-moment estimator $\xi_{Q,2}$ (see 
{next section). 
Fig.~\ref{fig:xi}(a) shows the LR estimators for $\xi$ and $\xi_Q$, comparing experiment and numerical simulations, and exposing the large difference between the two length scales. 
\begin{figure}
    \includegraphics[width=0.9\columnwidth]{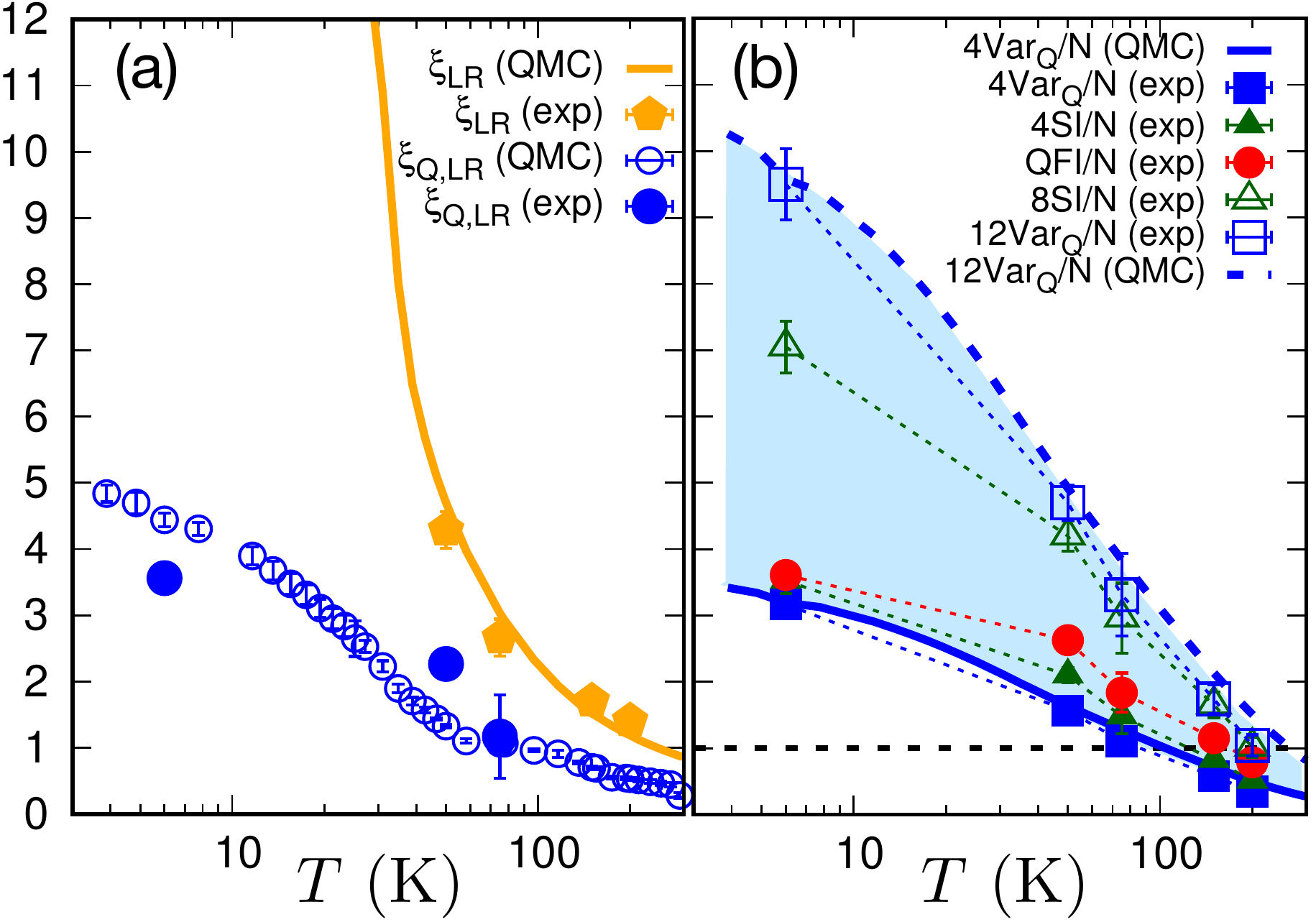}
    \caption{\label{fig:xi} {Quantum coherence length vs. entanglement depth.} (a) Quantum coherence length $\xi_Q$ vs. total correlation length $\xi$ as a function of temperature; (b) Entanglement depth estimated via the quantum fluctuations of the in-chain staggered magnetization, namely $4{\rm Var}_Q(M_{s,\rm chain})/N$ and  ${\rm QFI}(M_{s,\rm chain})/N$. The plot also shows $12{\rm Var}_Q/N$, which is an upper bound to ${\rm QFI}(M_{s,\rm chain})/N$, as well as the lower and upper bounds on ${\rm QFI}(M_{s,\rm chain})/N$ from the skew information, $4\rm SI/N$ and $8\rm SI/N$. Both panels compare QMC data and experimental data on KCuF$_3$. The error bars represent one standard deviation uncertainty.}
\end{figure}

One may wonder how the spatial extension of quantum correlations relates to multipartite entanglement, namely the entanglement depth. 
In fact there is a rigorous relationship: the entanglement depth along the chains is bounded from below by the Var$_Q$ density or the QFI density for the staggered magnetization  $M_s = \sum_r (-1)^r S_i^z$ of the individual chains, namely  ${\rm Var}_Q(M_{s,\rm chain})/N = \sum_r (-1)^r {\rm Cov}_Q(i,i+r)$ and  ${\rm QFI}(M_{s,\rm chain})/N = \sum_r (-1)^r {\rm QFIM}(i,i+r)$, 
where $r$ runs over distances along the chains. Indeed, when $4{\rm Var}_Q/N > k$ or when ${\rm QFI}/N > k$, one can conclude that spins in each chain exhibit at least $(k+1)$-partite entanglement \cite{Hyllus2012,Toth2012,FrerotR2016}. 
Fig.~\ref{fig:xi}(b) shows the temperature dependence of 4Var$_Q$ and QFI for KCuF$_3$, compared with the theoretical results for 4Var$_Q$. Interestingly, the entanglement depth estimate offered by these quantities is comparable to the quantum coherence length, rising up to $k+1 = 4$. In general one should expect quantum correlations to be systematically longer-ranged than the depth of entanglement, given that entanglement is a stronger form of correlation, and a state can be quantum correlated without being entangled \cite{AdessoBC2016}.

\begin{figure}
    \includegraphics[width=\columnwidth]{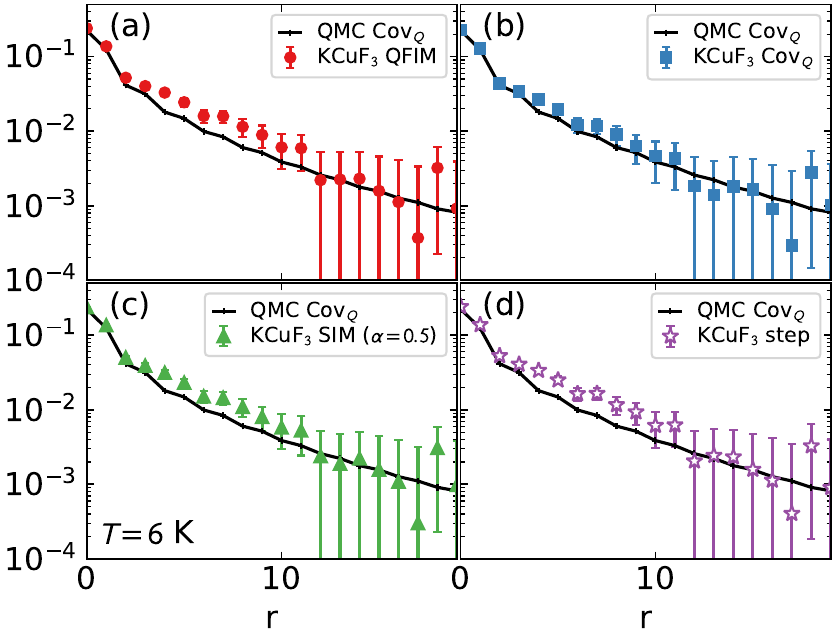}
    \caption{\label{fig:CorrelatorComparison} {Various quantum correlation functions evaluated for KCuF$_3$ at 6~K.} (a) The quantum Fisher information matrix {
    [Eq. \eqref{eq:QFIM}]}, scaled as QFIM/4. (b) The quantum covariance Cov$_Q$ {
    [Eq. \eqref{eq:varQ}]}. (c) The skew information matrix SIM {
    [Eq. \eqref{eq:SIM}]}. (d) Correlation function using the simple filter $h^\mathrm{step}(x)$ {
    [Eq. \eqref{eq:step}]}.  In all panels the experimental data are compared with the QMC data for ${\rm Cov}_Q$ for reference. The error bars represent one standard deviation uncertainty.}
\end{figure}

The various quantum correlation functions offered by Eq.~\eqref{eq:C} raise the question whether the quantum coherence length is uniquely defined, or whether it depends on the quantum filter $h$. 
Fig.~\ref{fig:CorrelatorComparison} shows that the same exponential decay is exhibited by all quantum correlation functions 
listed above (${\rm Cov}_Q$, QFIM, SIM). 
In fact, it is a rather robust feature, uniquely stemming from the high-pass nature of $h(x)$. 
To emphasize the universality, we also calculate the most naive correlation {
$C[O_i,O_j;h^{\rm step},\rho]$} using a step function filter: 
\begin{equation}
    h^{\rm step}(x) = \begin{cases}
        1 & \text{if } x/2 \geq 1\\
        0 & \text{if } x/2 < 1
    \end{cases}
    \label{eq:step}
\end{equation}
{
in Eq.~\eqref{eq:C}, }
such that all intensity below $\hbar \omega / 2 k_B T$ is suppressed. 
Although this filter function lacks the linear behavior at small $x$ required for a proper quantum-coherence measure, 
the plot in Fig. \ref{fig:CorrelatorComparison}(d) shows the same general behavior as the other quantum correlators. 
(Furthermore, the step function resembles filter functions that are naturally applied in neutron-scattering experiments, see discussion below.) 
Thus, although details of the quantum correlators depend on the filter function, the revealed length scale appears universal. 
The temperature dependence of all four quantum correlators is shown in Appendix~\ref{app:Comparing}.

\begin{figure*}
    \includegraphics[width=0.9\textwidth]{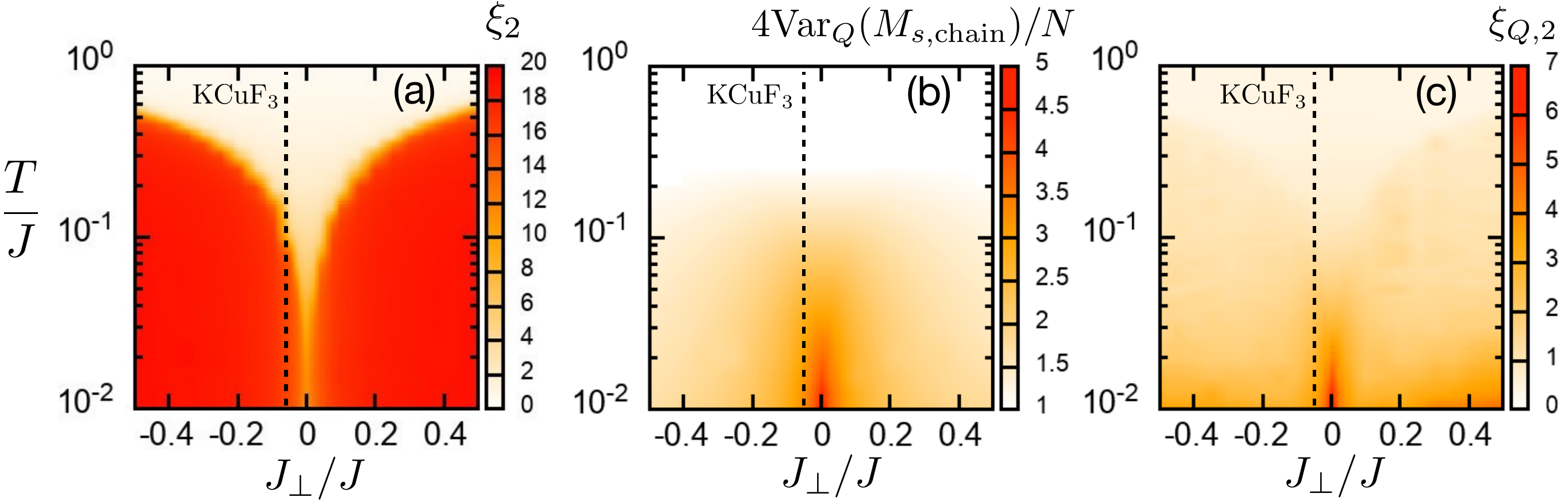}
    \caption{\label{fig:QMCcorrelation} {Total vs. quantum correlations for coupled Heisenberg chains from QMC data.} (a) Second-moment estimator for the in-chain correlation length $\xi_2$, showing clearly the N\'eel order; (b) Quantum variance per spin of the in-chain staggered magnetization, $4{\rm Var}_Q(M_s)/N$; (c) second-moment estimator for the in-chain quantum coherence length $\xi_{Q,2}$. 
    In all panels, the dashed line marks the value of $J_{\perp}/J$ realized by KCuF$_3$.}
\end{figure*}

\section{Redistribution of quantum correlations upon changing the interchain couplings}
We now embed the quantum correlations in KCuF$_3$ within the broader family of coupled spin chain models described by the Hamiltonian Eq.~\eqref{e.H}. 
{Using QMC simulations, we calculate correlations and quantum coherence with varying interchain coupling $J_{\perp}$, in order to explore the effect of the dimensionality of interactions.}
In the case of total correlations, an increase of $|J_{\perp}|$ at fixed temperature drives the system from quasi-one-dimensional magnetism towards three-dimensional magnetism, \emph{i.e.} towards a regime exhibiting stronger correlations in \emph{all} spatial directions, both transverse and longitudinal to the chains. 
This behavior is clearly exhibited by the {second-moment estimator for the} in-chain correlation length 
\begin{equation}
\xi_2^2 = \frac{1}{2} \frac{\sum_{r} r^2 |\langle \bm S_{i} \cdot \bm S_{i+r} \rangle|}{ \sum_{r}  |\langle \bm S_{i}\cdot \bm S_{i+r} \rangle|}   \label{eq:estimator}
\end{equation}
which allows for a systematic extraction of a typical length from all the correlation data produced with QMC across the vast parameter range explored in Fig. \ref{fig:QMCcorrelation}.
As shown in Fig.~\ref{fig:QMCcorrelation}(a), the total second moment estimator's sharp rise upon lowering the temperature marks the evolution of the N\'eel temperature with the interchain coupling \cite{Yasudaetal2005}.

On the other hand, quantum correlations are found to undergo a rather different fate along the dimensional crossover of the couplings. Fig.~\ref{fig:QMCcorrelation}(b-c) shows the $T$ and $J_{\perp}$ dependence of $\xi_{Q,2}$ 
(defined analogously to Eq.~\eqref{eq:estimator} by substituting 
$|\langle \bm S_{i} \cdot \bm S_{i+r} \rangle|$ with 
${\rm Cov}_Q(i,i+r)$) 
and ${\rm Var}_Q(M_{s,\rm chain})/N$ (a lower bound to the in-chain entanglement depth).   
Contrary to total correlations, quantum correlations along the chains appear to \emph{decrease} upon increasing the interchain couplings at low $T$ (and the N\'eel transition is nearly invisible to ${\rm Var}_Q(M_{s,\rm chain})/N$  -- see Ref.~\cite{Frerotetal2022} for further details on this aspect). The quantum coherence length $\xi_Q$ is in fact found to increase again at sufficiently low temperature and sufficiently strong $J_{\perp}$, but this is an effect driven by the appearance of thin tails in the quantum correlation function which have little effect on the integral given by the quantum variance (Appendix \ref{app:Interchain}). 


This result suggests that low-temperature quantum correlations in coupled-chain systems exhibit a form of \emph{monogamy} \cite{PhysRevA.61.052306}, since a dimensional crossover in the couplings entails their spatial redistribution from in-chain to inter-chain correlations. This result is quite insightful. In general, quantum correlations are not  {
necessarily} monogamous, as they can {
also be associated with states possessing} multipartite entanglement, which can imply an arbitrary number of degrees of freedom. The above behavior suggests that the Heisenberg two-spin couplings are primarily promoting quantum correlations in the form of two-spin entanglement --- presumably via singlet/triplet formation for antiferromagnetic/ferromagnetic couplings, respectively ---  which is indeed monogamous \cite{PhysRevA.61.052306, PhysRevLett.96.220503}. As a result, quantum correlations along the chains decrease over shorter length scales when the interchain coupling is increased. 
This result shows that, among the family of coupled-chain Heisenberg models, quasi-one-dimensional compounds such as KCuF$_3$ exhibit the strongest quantum correlations at short distance, whose detection via neutron scattering is most efficient. (In Appendix \ref{app:Interchain} we also show that interchain quantum correlations do not rise to the same strength as that of intrachain ones over the range of interchain couplings explored in this study.)

\section{Discussion}
These results have exciting implications far beyond 1D spin chains. 
Firstly, our results on 1D chains demonstrate that the spatial structure of quantum correlations reveals new \emph{quantitative} information about the dimensionality of quantum materials, a fundamental property inherently linked to quantum statistics and novel phases of matter. This is important as ``low-dimensional materials'' often exist in a three-dimensional host crystal and retain weak three-dimensional coupling. Our methods give access to the effective dimensionalities of both quantum and total correlations, which may be rather different, as our results clearly show. 
In this respect, it is important to note that our quantum correlator analysis is not restricted to neutron spectroscopy: any momentum-resolved probe of dynamic susceptibility associated with local Hermitian operators will work in the same fashion. 
For instance, quantum correlations in the charge sector could be probed via X-ray scattering \cite{schulke2007electron} or electron energy-loss spectroscopy \cite{SciPostPhys.3.4.026}, offering complementary pictures of quantum coherence in a huge variety of quantum materials.

Secondly, the quantum correlators are model-independent, which allows precise statements to be made about materials even in the absence of a tractable theory. Therefore they may yield important information about enigmatic condensed matter states. 
For example, one could evaluate how the spatial structure of quantum correlations changes across quantum phase transitions (as in, e.g., heavy fermion materials \cite{gegenwart2008quantum} or quantum magnets under fields). Recent works \cite{Frerot2019,Gabbriellietal2018} showed that quantum correlations can reconstruct the quantum critical fan occurring at finite temperatures above such quantum critical points, thus certifying quantum criticality and delineating the range of genuine quantum critical behavior. 
Within the space of coupled spin chains, it would be interesting to apply the same analysis to systems with frustrated interchain coupling, such as Cs$_2$CuCl$_4$ \cite{PhysRevLett.88.137203, PhysRevB.82.014421}, in order to test if frustration can stabilize the intrachain quantum {
coherence} length compared to the unfrustrated case studied here. 

Thirdly and more generally, our results advance the synthesis of condensed matter physics and quantum information. 
Specifically, we show that experimental momentum-resolved dynamical response functions at thermal equilibrium can be mined for a wealth of many-body quantum information. 
The ability to do this for a thermodynamic system at a well-defined temperature is not shared by many other platforms for quantum many-body physics. (For example, most quantum many-body physics simulators 
based on atomic physics platforms do not operate at thermal equilibrium; or, if they do so, their temperature is not easily accessible or cannot easily be held fixed \cite{McKay2011,Trotzky2010,Carcy2021}. As a consequence, a similar analysis to ours cannot be straightforwardly conducted with e.g. cold atoms.) Hence our results indicate a clear path for experiments on quantum materials to positively contribute to quantum information theory, by revealing the microscopic structure of quantum correlations in many-body states. 

{
We also note that quantum correlations can be extracted without {
Fourier-transforming the momentum-resolved spectroscopic data}. If the integral {
over frequencies defining quantum correlations} in Eq. \eqref{eq:C} is carried out  {
using  $\chi''(\bf Q,\omega)$, one can extract a quantum structure factor, and hence a quantum coherence length by fitting the structure factor to} a resolution-convolved Lorentzian---in an analogous way to how total correlations are conventionally extracted from energy-integrated Bragg {
peaks}. Although much information about the detailed spatial-dependence {
of quantum correlations} is lost {
using this procedure}, it may prove an easier experimental way to evaluate a quantum {
coherence} length in higher-dimensional materials. 
}

On a different note, the fact that the step function filter captures the same behavior as the other quantum correlators suggests that approximate results for the quantum length scale can be experimentally obtained by neutron diffraction methods. At low temperatures where $\chi''(\mathbf{k},\omega) \approx \pi S(\mathbf{k},\omega)$, the filter function $h^\mathrm{step}(x)$ can be traded for a physical neutron transmission filter (e.g., beryllium powder) \cite{egelstaff1954design,tennant1988performance}, tuning the incident neutron energy to act as a high-pass filter {
in energy transfer $\hbar \omega$ (absorbing neutrons with large final energy)}. For suitable systems (specifically, low-bandwidth materials such as CuSO$_4\cdot$5D$_2$O \cite{Mourigal2013}, YbAlO$_3$ \cite{Wu2019}, and Cs$_2$CoCl$_4$ \cite{PhysRevLett.127.037201}) it offers a path towards quickly identifying if a material has significant quantum correlations.

\section{Conclusions} 
We have shown how the spatial structure of quantum correlation functions for quantum spin systems can be extracted from neutron spectroscopy data. The data reveal the existence of a fundamental length scale of quantum mechanical origin --- the quantum coherence length --- limiting the range of quantum correlations at all finite temperatures, which is wildly different from the correlation length. Our study also highlights the role of dimensionality on quantum correlations, showing that a stronger coupling between Heisenberg chains leads to a redistribution of quantum correlations from the chains to the transverse directions, in contrast to total correlations. 
As a consequence, within the family of coupled-chain compounds, systems close to the one-dimensional limit --- such as KCuF$_3$ --- exhibit the strongest short-range quantum correlations and the weakest total correlations. 
The fact that quantum correlators enabled new observations---even for a well-studied model as the one-dimensional Heisenberg chain---indicates that quantum correlator analysis could be a powerful new way of assessing the underlying quantum state of a vast number of quantum materials, both low- and higher-dimensional.


\begin{acknowledgments}
We thank I. Fr\'erot for valuable discussions. 
The work by A.S. and D.A.T. is supported by the Quantum Science Center (QSC), a National Quantum Information Science Research Center of the U.S. Department of Energy (DOE). 
The work of P.L. and E.D. was supported by the U.S. Department of Energy, Office of Science, Basic Energy Sciences, Materials Sciences and Engineering Division. 
All QMC calculations have been performed on the PSMN cluster at the ENS of Lyon. 
\end{acknowledgments}


\appendix

\section{Experimental data processing \label{app:ExperimentalData}}

{
The explicit data analysis protocol for extracting real space quantum correlators is as follows:
\begin{enumerate}
    \item Isolate magnetic scattering in a full Brillouin zone.
    \item Correct for the form factor and $g$-factor.
    \item Correct for the polarization factor (if anisotropic exchange exists).
    \item Normalize data to absolute units and convert to $\chi''$  \cite{Xu_AbsUnits_2013}.
    \item Take Fourier transform from reciprocal to real space.
    \item Apply filter function $h$.
    \item Analyze spatial dependence.
\end{enumerate}
Note that steps 5 and 6 can be exchanged for equivalent results. Note also that step 3 is only necessary if anisotropy is present such that $S_{xx} \neq S_{yy} \neq S_{zz}$, in which case one must correct for the experimental neutron polarization factor \cite{Squires}. This can be done with theoretical modeling \cite{PhysRevLett.127.037201}, or more generally by measuring polarized neutron scattering.  
In this study, we used a highly isotropic system and thus step 3 was not necessary. 
}

All experimental data used in this study was previously published in \cite{Scheie2022}, which involved a composite data set from two different neutron experiments for temperatures $T=75$~K and above: SEQUOIA at the SNS \cite{Granroth_2010} for low energy, and MAPS at ISIS \cite{perring1994proceedings} for high energy. The data at $T=6$~K and 50~K are from ISIS. 
For the quantum correlators, we applied the formulae to the data as previously processed. However, for the conventional correlation length at $T=6$~K, we applied a resolution deconvolution. This is because KCuF$_3$ at $T=6$~K has long range magnetic order and a very long correlation length ($\sim 700$~sites \cite{ikeda1973neutron}). Consequently, resolution broadening has to be corrected before the true long range correlations will emerge. 
We did this by fitting the $q=\pi$, $T=6$~K\, $\hbar \omega = 0$  scattering to a Gaussian function, and dividing the Fourier transformed structure factor by the Fourier transformed Gaussian function. This resulted in a visible increase in total correlations for $r>15$. For only the $T=6$~K conventional correlations did this correction make any visible difference.

It should be noted that the experimental energy and momentum resolution limit how low in temperature and how far in real space one can analyze the quantum correlations. As the resolution of each improves, one can evaluate lower temperatures and larger spatial distances. 
Although the limits of each is not something we explore in this manuscript, one can still use resolution as a rough guideline: 
if $\Delta \hbar \omega$ is the energy resolution,  one {
can only be sensitive to temperatures such that $k_B T > \Delta \hbar \omega$, and one cannot fully appreciate the enhancement of quantum correlations when cooling below this temperature scale}. Similarly, if  $\Delta Q$ is the average momentum resolution along a particular reciprocal space direction, one can only evaluate distances $\leq \frac{2 \pi}{\Delta Q}$ along that direction in real space. 

\section{Evolution of quantum correlations upon changing the interchain couplings \label{app:Interchain}}

Here we discuss the detailed evolution of the spatial structure of quantum correlations upon changing the strength of the coupling between Heisenberg chains, as stemming from our experimental as well as theoretical data.  Fig.~\ref{fig:coupled_vs_1d} shows the comparison between the experimental data on two different quantum correlation functions (quantum covariance and the quantum Fisher information matrix) for KCuF$_3$ at $T = 6$K, and theoretical data obtained for a single one-dimensional chain. In particular the theory data on quantum covariance are obtained via quantum Monte Carlo (QMC) as in the main text. On the other hand, the data for the quantum Fisher information matrix are inaccessible to QMC, because they require the full knowledge of the dynamical susceptibility. For one-dimensional systems, this knowledge can be obtained using the density-matrix renormalization group (DMRG) \cite{PhysRevLett.69.2863, PhysRevB.48.10345}, which allows for the calculation of $S(k,\omega)$ at finite temperature \cite{PhysRevB.72.220401, PhysRevB.79.245101}. Here we extend the finite-$T$ calculations reported in Ref.~\cite{PhysRevB.103.224434} down to $T=6$~K. We use the DMRG++ software \cite{Alvarez2009} to study a system with open boundary conditions, consisting of $L=50$ physical sites, and $50$ ``ancilla'' sites. $S(k,\omega)$ spectra are calculated using the Krylov-space correction vector method \cite{PhysRevB.60.335, PhysRevE.94.053308}, with a Lorentzian energy broadening with half width at half maximum $\eta=0.1J$. For details on how to reproduce the DMRG calculations, see the supplemental material of Ref.~\cite{PhysRevB.103.224434}.

As seen in Fig.~\ref{fig:coupled_vs_1d}, the predictions for the quantum correlation functions of a single one-dimensional chain lie systematically above the measured values for KCuF$_3$, while a much better quantitative agreement is obtained when taking into account the small albeit finite interchain coupling $J_{\perp}$, as shown in Fig.~\ref{fig:coupled_vs_1d}(b).  This indicates that, 1) the resolution of the experiment is clearly sufficient to reconstruct the difference between isolated vs. weakly coupled Heisenberg chains, and 2) moving from a single chain to coupled chains, quantum correlations reorganize spatially, in such a way that correlations along the chains are suppressed.

\begin{figure}
    \includegraphics[width=0.48\textwidth]{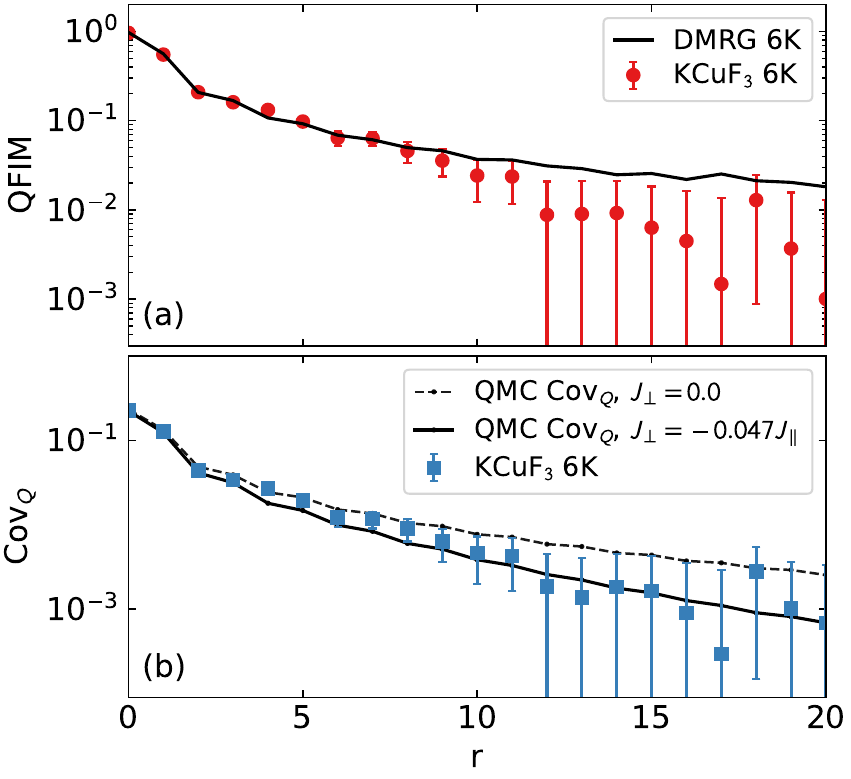}
    \caption{\label{fig:coupled_vs_1d}{Comparison between 1d theoretical QFIM (a) and quantum covariance (b), and experimental KCuF$_3$ data at 6K.} For comparison, the QMC result for KCuF$_3$ $J_\perp$ is also shown in (b). Note the $r>10$ experimental values are systematically smaller than the theoretical 1D calculations as a consequence of finite interchain coupling $J_{\perp}$.}
\end{figure}

\begin{figure*}
    \includegraphics[width=0.9\textwidth]{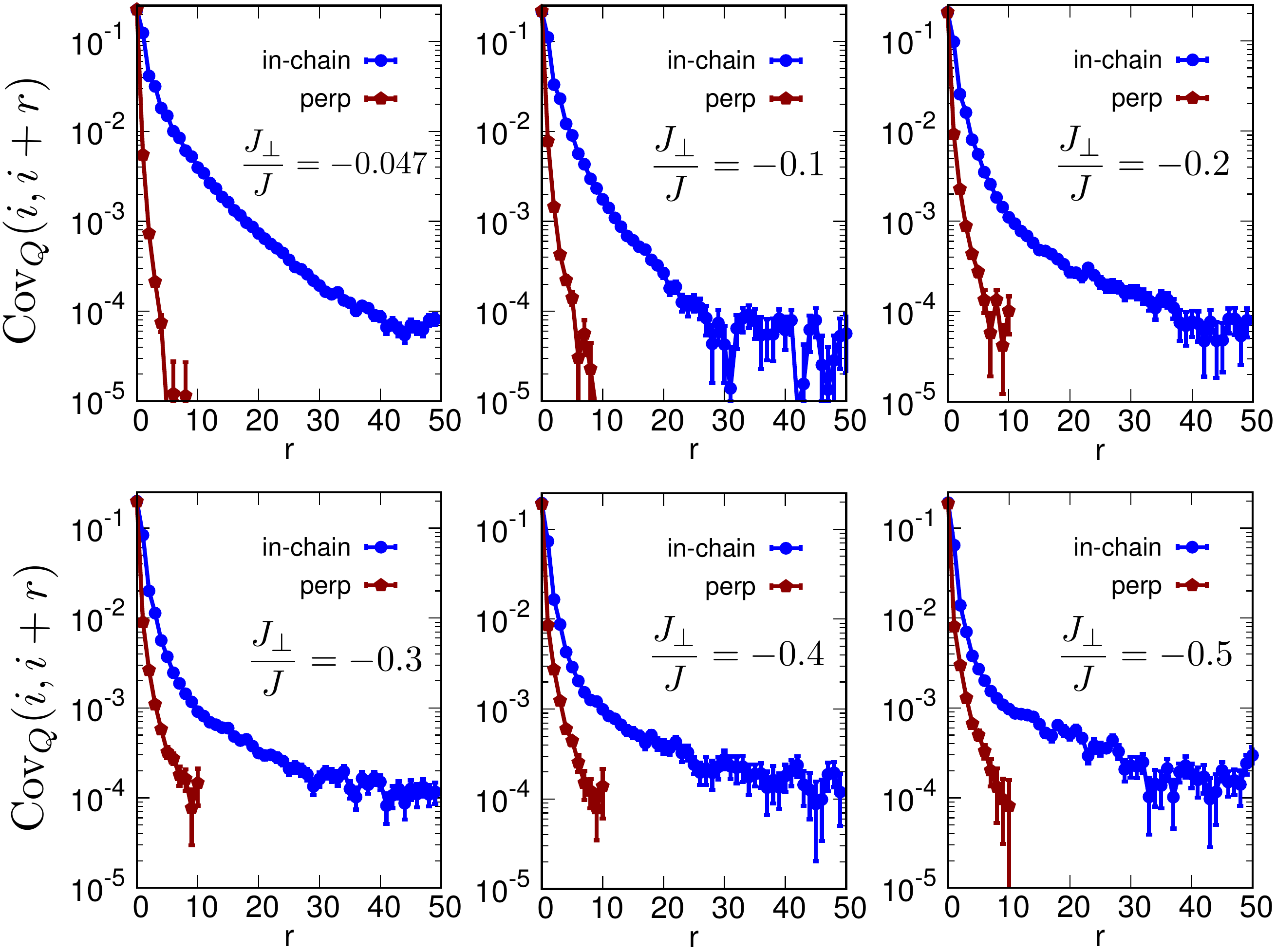}
    \caption{\label{fig:perpcorr} {Dependence of Cov$_Q$ on $J_\perp$.} Dependence of the quantum covariance on the ferromagnetic interchain coupling $J_{\perp}$ for $J_\perp/J = -0.047$ (as for KCuF$_3$), $-0.1$, $-0.2$, $-0.3$, $-0.4$ and $-0.5$, at a temperature $T/k_B = 6J/388$ (corresponding to 6K for KCuF$_3$).  The panels show QMC data obtained for a system with size $100\times 20 \times 20$. Each panel shows the in-chain correlations (${\rm Cov}_Q(i,i+x)$, taking $x$ as the lattice direction parallel to the chains) and the correlations perpendicular to the chain (e.g. ${\rm Cov}_Q(i,i+y)$).}
\end{figure*}

We examine this trend systematically via QMC, by monitoring how the spatial structure of the quantum covariance changes upon increasing the magnitude of the ferromagnetic coupling ($J_{\perp} < 0$) between the chains. In particular we examine the quantum covariance at $T=6$~K (assuming an in-chain coupling equal to that of KCuF$_3$) along the chains, namely ${\rm Cov}_Q(i,i+x)$, taking x as the direction of extension of the chains; and perpendicular to the chains, namely ${\rm Cov}_Q(i,i+y)$, where $y$ is one of the two perpendicular lattice directions. 
Fig.~\ref{fig:perpcorr} shows that, upon increasing $|J_{\perp}|$, the correlations along a coordinate axis perpendicular to the chain become stronger, as can be trivially expected -- albeit remaining much weaker than the in-chain correlations for the whole range of values of $J_{\perp}$ we explored ($|J_{\perp}|\leq J/2$). On the other hand, the in-chain quantum covariance undergoes a much more complex evolution: it becomes significantly weaker at short range when $|J_{\perp}|$ increases, witnessing a form of monogamy of short-range quantum correlations, as discussed in the main text. Yet the behavior at long range shows an opposite trend for sufficiently large $|J_{\perp}|$,  as the in-chain quantum covariance develops a stronger tail. This tail can be associated with the appearance of long-range \emph{multipartite} quantum correlations. Such correlations are expected in a long-range-ordered quantum ground state, such as that of a system of coupled Heisenberg chains; and their multipartite nature makes them no longer monogamous. Nonetheless, the long-range tail is rather thin, and it gives a small contribution to the quantum variance of the in-chain staggered magnetization, so that the global trend is a decrease of this quantity with $|J_{\perp}|$, as shown in Fig. \ref{fig:QMCcorrelation}(b) of the main text. 

\begin{figure}
    \includegraphics[width=0.8\columnwidth]{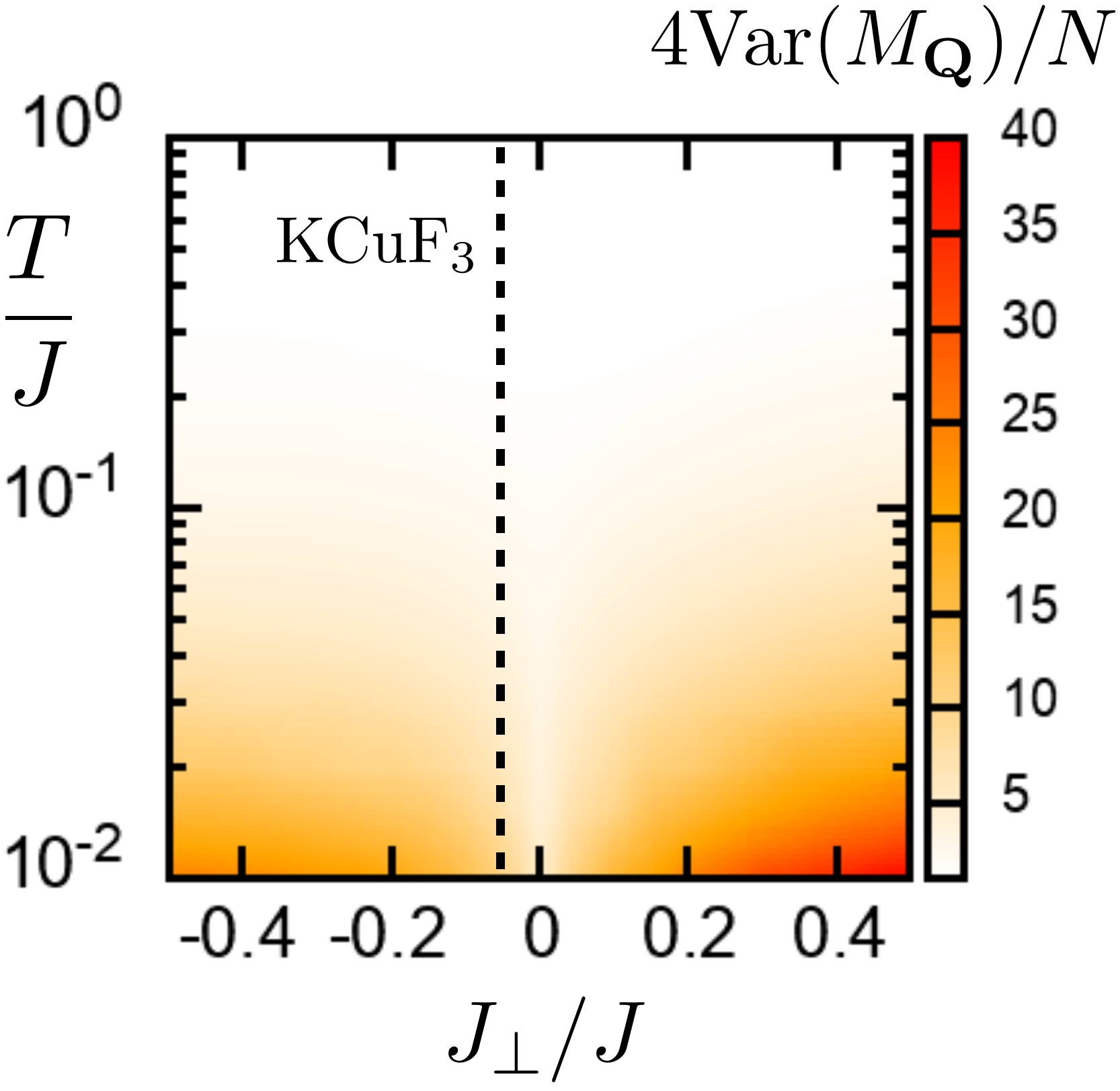}
    \caption{\label{fig:qvariance} {Quantum variance density of the order parameter for coupled Heisenberg chains.} The ordering vector ${\bf Q}$ is $(\pi,0,0)$ for $J_{\perp}<0$ and $(\pi,\pi,\pi)$ for $J_{\perp}=0$.}
\end{figure}

The buildup of increasingly strong multipartite quantum correlations and entanglement upon coupling the chains is clearly exhibited in Fig.~\ref{fig:qvariance}, in which the quantum variance density of the order parameter $M_{\bf Q} = \sum_i e^{i\bf Q \cdot {\bf r}_i} S_i^z$ is shown; the ordering vector ${\bf Q}$ being $(\pi,0,0)$ for $J_{\perp}<0$ and $(\pi,\pi,\pi)$ for $J_{\perp}=0$. One clearly observes that rather massive entanglement sets in at low temperatures in the more strongly coupled chains -- involving $> 40$ spins within the temperature and parameter range we explored. Yet this behavior really stems from the buildup of correlations transverse to the chains -- as can be easily deduced by comparing with Fig. \ref{fig:QMCcorrelation}(b) of the main text.

%

\section{Comparing quantum correlators \label{app:Comparing}}

\begin{figure*}
    \includegraphics[width=0.85\textwidth]{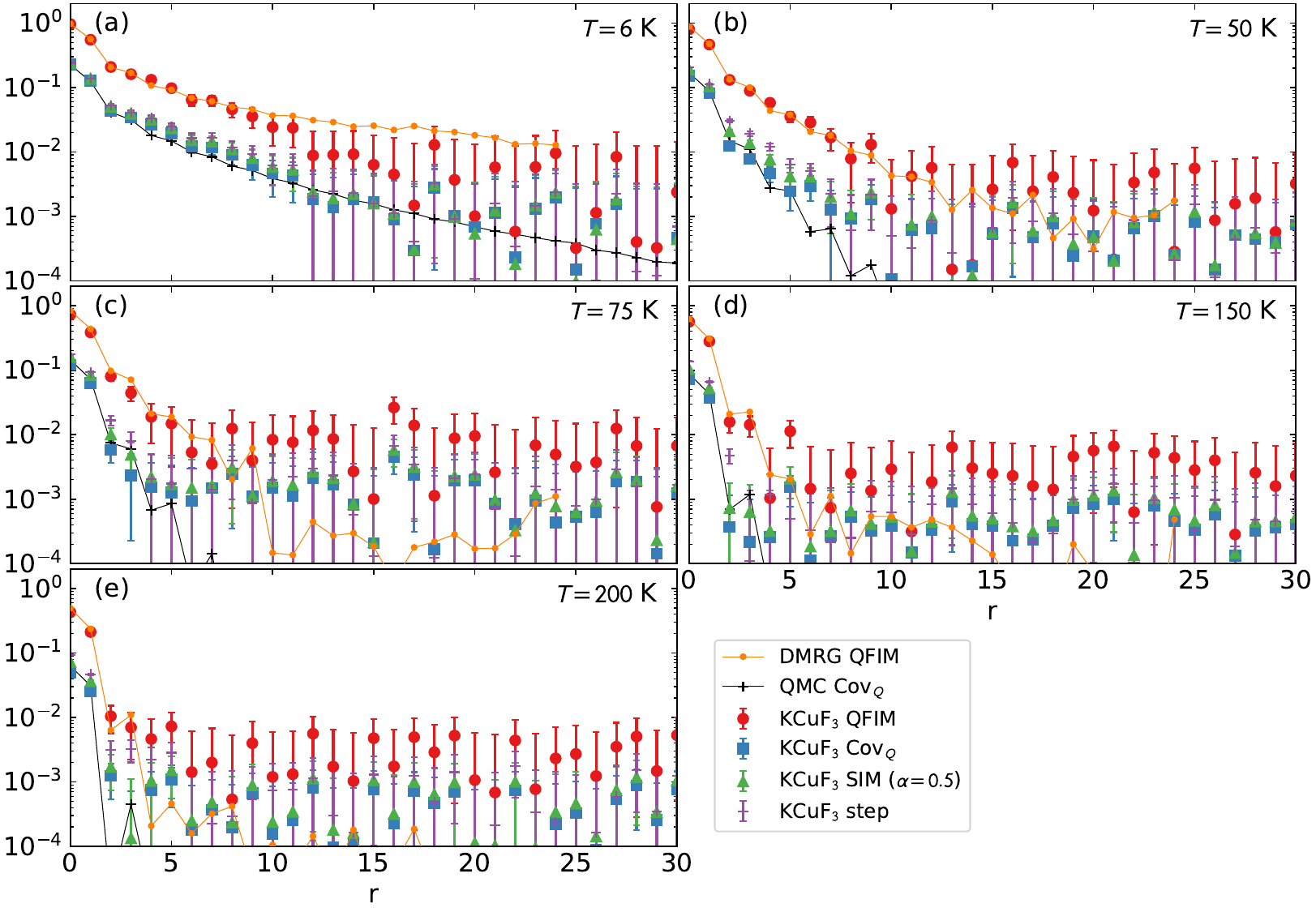}
    \caption{\label{fig:CorrelatorComparisonSI}{Full temperature dependence of the quantum correlation functions.} Absolute values of various definitions of quantum correlator applied to KCuF$_3$ at temperatures between $T=6$~K and $T=200$~K, as well as QMC and DMRG calculations for comparison. Panel (a) corresponds to Fig. \ref{fig:CorrelatorComparison} in the main text, but with the raw values of QFI matrix.}
\end{figure*}

\begin{figure}
    \includegraphics[width=0.95\columnwidth]{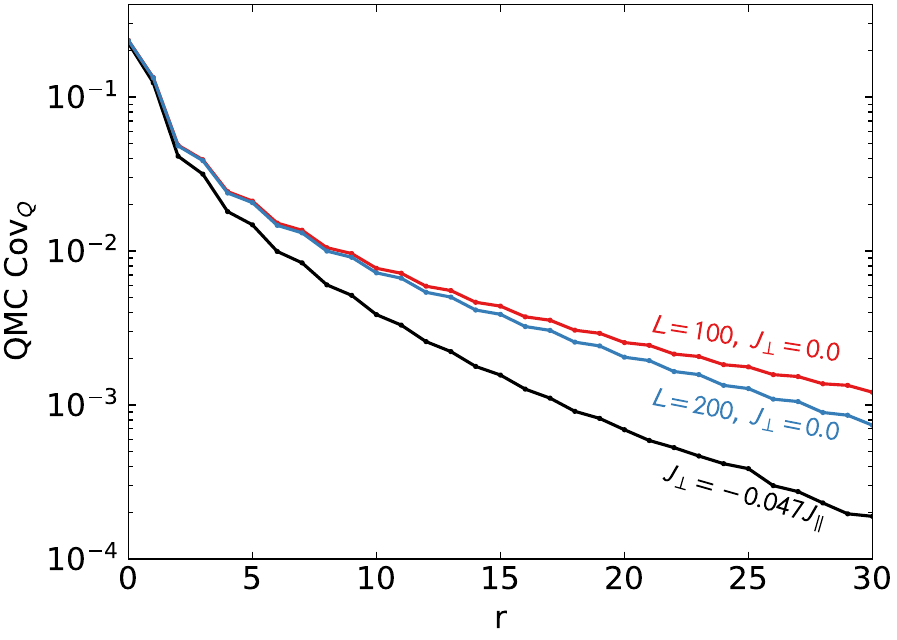}
    \caption{\label{fig:QMC_length}{Quantum covariance from quantum Monte Carlo (QMC) for coupled and uncoupled chains.} The quantum covariance is much longer ranged in the one-dimensional $J_{\perp} = 0$ limit. Meanwhile the chain length slightly decreases the length scale, in accord with the periodic boundary being further away.}
\end{figure}

For a comparison of the experimental and theoretical quantum correlators at all measured temperatures, see Fig. \ref{fig:CorrelatorComparisonSI}. Note that above $T=75$~K, only a few nearest neighbors have quantum correlations distinguishable from the large $r$ background noise. This is a consequence of the experimental background in KCuF$_3$, which was previously discussed in Ref.~\cite{PhysRevB.103.224434}.

The DMRG data for the QFIM on isolated chains at finite temperature, shown in Fig.~\ref{fig:CorrelatorComparisonSI}, may erroneously suggest that the latter quantity possesses a decay length which is systematically larger than that of e.g. Cov$_Q$. This discrepancy is in fact not intrinsic, but it is rather the result of the fact that the DMRG calculations are done on isolated chains, while the theoretical curves for Cov$_Q$ are QMC data for coupled chains. Indeed, as noted in the main text, the strength of the interchain coupling $J_{\perp}$ makes a dramatic difference in the length scale of the quantum correlator. Figure \ref{fig:QMC_length} shows this more explicitly, with ${\rm Cov}_Q$ plotted for the KCuF$_3$ value of $J_{\perp}$ and for $J_{\perp} = 0$. See also the detailed discussion of this topic in Appendix~\ref{app:Interchain}.

In the case of one-dimensional QMC simulations, we also show the comparison between the quantum covariances on two different system sizes ($L=100$ and $L=200$). The data display minor differences over the range of distances which are relevant for the experiment. We therefore conclude that the QMC data we use for the quantum covariance are essentially devoid of significant finite-size effects.


\begin{thebibliography}{71}%
\makeatletter
\providecommand \@ifxundefined [1]{%
 \@ifx{#1\undefined}
}%
\providecommand \@ifnum [1]{%
 \ifnum #1\expandafter \@firstoftwo
 \else \expandafter \@secondoftwo
 \fi
}%
\providecommand \@ifx [1]{%
 \ifx #1\expandafter \@firstoftwo
 \else \expandafter \@secondoftwo
 \fi
}%
\providecommand \natexlab [1]{#1}%
\providecommand \enquote  [1]{``#1''}%
\providecommand \bibnamefont  [1]{#1}%
\providecommand \bibfnamefont [1]{#1}%
\providecommand \citenamefont [1]{#1}%
\providecommand \href@noop [0]{\@secondoftwo}%
\providecommand \href [0]{\begingroup \@sanitize@url \@href}%
\providecommand \@href[1]{\@@startlink{#1}\@@href}%
\providecommand \@@href[1]{\endgroup#1\@@endlink}%
\providecommand \@sanitize@url [0]{\catcode `\\12\catcode `\$12\catcode
  `\&12\catcode `\#12\catcode `\^12\catcode `\_12\catcode `\%12\relax}%
\providecommand \@@startlink[1]{}%
\providecommand \@@endlink[0]{}%
\providecommand \url  [0]{\begingroup\@sanitize@url \@url }%
\providecommand \@url [1]{\endgroup\@href {#1}{\urlprefix }}%
\providecommand \urlprefix  [0]{URL }%
\providecommand \Eprint [0]{\href }%
\providecommand \doibase [0]{https://doi.org/}%
\providecommand \selectlanguage [0]{\@gobble}%
\providecommand \bibinfo  [0]{\@secondoftwo}%
\providecommand \bibfield  [0]{\@secondoftwo}%
\providecommand \translation [1]{[#1]}%
\providecommand \BibitemOpen [0]{}%
\providecommand \bibitemStop [0]{}%
\providecommand \bibitemNoStop [0]{.\EOS\space}%
\providecommand \EOS [0]{\spacefactor3000\relax}%
\providecommand \BibitemShut  [1]{\csname bibitem#1\endcsname}%
\let\auto@bib@innerbib\@empty
\bibitem [{\citenamefont {Horodecki}\ \emph {et~al.}(2009)\citenamefont
  {Horodecki}, \citenamefont {Horodecki}, \citenamefont {Horodecki},\ and\
  \citenamefont {Horodecki}}]{RevModPhys.81.865}%
  \BibitemOpen
  \bibfield  {author} {\bibinfo {author} {\bibfnamefont {R.}~\bibnamefont
  {Horodecki}}, \bibinfo {author} {\bibfnamefont {P.}~\bibnamefont
  {Horodecki}}, \bibinfo {author} {\bibfnamefont {M.}~\bibnamefont
  {Horodecki}},\ and\ \bibinfo {author} {\bibfnamefont {K.}~\bibnamefont
  {Horodecki}},\ }\bibfield  {title} {\bibinfo {title} {Quantum entanglement},\
  }\href {https://doi.org/10.1103/RevModPhys.81.865} {\bibfield  {journal}
  {\bibinfo  {journal} {Rev. Mod. Phys.}\ }\textbf {\bibinfo {volume} {81}},\
  \bibinfo {pages} {865} (\bibinfo {year} {2009})}\BibitemShut {NoStop}%
\bibitem [{\citenamefont {Brunner}\ \emph {et~al.}(2014)\citenamefont
  {Brunner}, \citenamefont {Cavalcanti}, \citenamefont {Pironio}, \citenamefont
  {Scarani},\ and\ \citenamefont {Wehner}}]{Brunneretal2014}%
  \BibitemOpen
  \bibfield  {author} {\bibinfo {author} {\bibfnamefont {N.}~\bibnamefont
  {Brunner}}, \bibinfo {author} {\bibfnamefont {D.}~\bibnamefont {Cavalcanti}},
  \bibinfo {author} {\bibfnamefont {S.}~\bibnamefont {Pironio}}, \bibinfo
  {author} {\bibfnamefont {V.}~\bibnamefont {Scarani}},\ and\ \bibinfo {author}
  {\bibfnamefont {S.}~\bibnamefont {Wehner}},\ }\bibfield  {title} {\bibinfo
  {title} {Bell nonlocality},\ }\href
  {https://doi.org/10.1103/RevModPhys.86.419} {\bibfield  {journal} {\bibinfo
  {journal} {Rev. Mod. Phys.}\ }\textbf {\bibinfo {volume} {86}},\ \bibinfo
  {pages} {419} (\bibinfo {year} {2014})}\BibitemShut {NoStop}%
\bibitem [{\citenamefont {Wang}\ \emph {et~al.}(2016)\citenamefont {Wang},
  \citenamefont {Chen}, \citenamefont {Li}, \citenamefont {Huang},
  \citenamefont {Liu}, \citenamefont {Chen}, \citenamefont {Luo}, \citenamefont
  {Su}, \citenamefont {Wu}, \citenamefont {Li}, \citenamefont {Lu},
  \citenamefont {Hu}, \citenamefont {Jiang}, \citenamefont {Peng},
  \citenamefont {Li}, \citenamefont {Liu}, \citenamefont {Chen}, \citenamefont
  {Lu},\ and\ \citenamefont {Pan}}]{Wangetal2016}%
  \BibitemOpen
  \bibfield  {author} {\bibinfo {author} {\bibfnamefont {X.-L.}\ \bibnamefont
  {Wang}}, \bibinfo {author} {\bibfnamefont {L.-K.}\ \bibnamefont {Chen}},
  \bibinfo {author} {\bibfnamefont {W.}~\bibnamefont {Li}}, \bibinfo {author}
  {\bibfnamefont {H.-L.}\ \bibnamefont {Huang}}, \bibinfo {author}
  {\bibfnamefont {C.}~\bibnamefont {Liu}}, \bibinfo {author} {\bibfnamefont
  {C.}~\bibnamefont {Chen}}, \bibinfo {author} {\bibfnamefont {Y.-H.}\
  \bibnamefont {Luo}}, \bibinfo {author} {\bibfnamefont {Z.-E.}\ \bibnamefont
  {Su}}, \bibinfo {author} {\bibfnamefont {D.}~\bibnamefont {Wu}}, \bibinfo
  {author} {\bibfnamefont {Z.-D.}\ \bibnamefont {Li}}, \bibinfo {author}
  {\bibfnamefont {H.}~\bibnamefont {Lu}}, \bibinfo {author} {\bibfnamefont
  {Y.}~\bibnamefont {Hu}}, \bibinfo {author} {\bibfnamefont {X.}~\bibnamefont
  {Jiang}}, \bibinfo {author} {\bibfnamefont {C.-Z.}\ \bibnamefont {Peng}},
  \bibinfo {author} {\bibfnamefont {L.}~\bibnamefont {Li}}, \bibinfo {author}
  {\bibfnamefont {N.-L.}\ \bibnamefont {Liu}}, \bibinfo {author} {\bibfnamefont
  {Y.-A.}\ \bibnamefont {Chen}}, \bibinfo {author} {\bibfnamefont {C.-Y.}\
  \bibnamefont {Lu}},\ and\ \bibinfo {author} {\bibfnamefont {J.-W.}\
  \bibnamefont {Pan}},\ }\bibfield  {title} {\bibinfo {title} {Experimental
  ten-photon entanglement},\ }\href
  {https://doi.org/10.1103/PhysRevLett.117.210502} {\bibfield  {journal}
  {\bibinfo  {journal} {Phys. Rev. Lett.}\ }\textbf {\bibinfo {volume} {117}},\
  \bibinfo {pages} {210502} (\bibinfo {year} {2016})}\BibitemShut {NoStop}%
\bibitem [{\citenamefont {Thomas}\ \emph {et~al.}(2022)\citenamefont {Thomas},
  \citenamefont {Ruscio}, \citenamefont {Morin},\ and\ \citenamefont
  {Rempe}}]{Thomasetal2022}%
  \BibitemOpen
  \bibfield  {author} {\bibinfo {author} {\bibfnamefont {P.}~\bibnamefont
  {Thomas}}, \bibinfo {author} {\bibfnamefont {L.}~\bibnamefont {Ruscio}},
  \bibinfo {author} {\bibfnamefont {O.}~\bibnamefont {Morin}},\ and\ \bibinfo
  {author} {\bibfnamefont {G.}~\bibnamefont {Rempe}},\ }\bibfield  {title}
  {\bibinfo {title} {Efficient generation of entangled multiphoton graph states
  from a single atom},\ }\href {https://doi.org/10.1038/s41586-022-04987-5}
  {\bibfield  {journal} {\bibinfo  {journal} {Nature}\ }\textbf {\bibinfo
  {volume} {608}},\ \bibinfo {pages} {677} (\bibinfo {year}
  {2022})}\BibitemShut {NoStop}%
\bibitem [{\citenamefont {Monz}\ \emph {et~al.}(2011)\citenamefont {Monz},
  \citenamefont {Schindler}, \citenamefont {Barreiro}, \citenamefont {Chwalla},
  \citenamefont {Nigg}, \citenamefont {Coish}, \citenamefont {Harlander},
  \citenamefont {H\"ansel}, \citenamefont {Hennrich},\ and\ \citenamefont
  {Blatt}}]{Monzetal2011}%
  \BibitemOpen
  \bibfield  {author} {\bibinfo {author} {\bibfnamefont {T.}~\bibnamefont
  {Monz}}, \bibinfo {author} {\bibfnamefont {P.}~\bibnamefont {Schindler}},
  \bibinfo {author} {\bibfnamefont {J.~T.}\ \bibnamefont {Barreiro}}, \bibinfo
  {author} {\bibfnamefont {M.}~\bibnamefont {Chwalla}}, \bibinfo {author}
  {\bibfnamefont {D.}~\bibnamefont {Nigg}}, \bibinfo {author} {\bibfnamefont
  {W.~A.}\ \bibnamefont {Coish}}, \bibinfo {author} {\bibfnamefont
  {M.}~\bibnamefont {Harlander}}, \bibinfo {author} {\bibfnamefont
  {W.}~\bibnamefont {H\"ansel}}, \bibinfo {author} {\bibfnamefont
  {M.}~\bibnamefont {Hennrich}},\ and\ \bibinfo {author} {\bibfnamefont
  {R.}~\bibnamefont {Blatt}},\ }\bibfield  {title} {\bibinfo {title} {14-qubit
  entanglement: Creation and coherence},\ }\href
  {https://doi.org/10.1103/PhysRevLett.106.130506} {\bibfield  {journal}
  {\bibinfo  {journal} {Phys. Rev. Lett.}\ }\textbf {\bibinfo {volume} {106}},\
  \bibinfo {pages} {130506} (\bibinfo {year} {2011})}\BibitemShut {NoStop}%
\bibitem [{\citenamefont {Schmied}\ \emph {et~al.}(2016)\citenamefont
  {Schmied}, \citenamefont {Bancal}, \citenamefont {Allard}, \citenamefont
  {Fadel}, \citenamefont {Scarani}, \citenamefont {Treutlein},\ and\
  \citenamefont {Sangouard}}]{Schmied16}%
  \BibitemOpen
  \bibfield  {author} {\bibinfo {author} {\bibfnamefont {R.}~\bibnamefont
  {Schmied}}, \bibinfo {author} {\bibfnamefont {J.-D.}\ \bibnamefont {Bancal}},
  \bibinfo {author} {\bibfnamefont {B.}~\bibnamefont {Allard}}, \bibinfo
  {author} {\bibfnamefont {M.}~\bibnamefont {Fadel}}, \bibinfo {author}
  {\bibfnamefont {V.}~\bibnamefont {Scarani}}, \bibinfo {author} {\bibfnamefont
  {P.}~\bibnamefont {Treutlein}},\ and\ \bibinfo {author} {\bibfnamefont
  {N.}~\bibnamefont {Sangouard}},\ }\bibfield  {title} {\bibinfo {title} {Bell
  correlations in a {Bose-Einstein} condensate},\ }\href
  {https://doi.org/10.1126/science.aad8665} {\bibfield  {journal} {\bibinfo
  {journal} {Science}\ }\textbf {\bibinfo {volume} {352}},\ \bibinfo {pages}
  {441} (\bibinfo {year} {2016})}\BibitemShut {NoStop}%
\bibitem [{\citenamefont {Omran}\ \emph {et~al.}(2019)\citenamefont {Omran},
  \citenamefont {Levine}, \citenamefont {Keesling}, \citenamefont {Semeghini},
  \citenamefont {Wang}, \citenamefont {Ebadi}, \citenamefont {Bernien},
  \citenamefont {Zibrov}, \citenamefont {Pichler}, \citenamefont {Choi},
  \citenamefont {Cui}, \citenamefont {Rossignolo}, \citenamefont {Rembold},
  \citenamefont {Montangero}, \citenamefont {Calarco}, \citenamefont {Endres},
  \citenamefont {Greiner}, \citenamefont {Vuleti\'c},\ and\ \citenamefont
  {Lukin}}]{Omranetal2019}%
  \BibitemOpen
  \bibfield  {author} {\bibinfo {author} {\bibfnamefont {A.}~\bibnamefont
  {Omran}}, \bibinfo {author} {\bibfnamefont {H.}~\bibnamefont {Levine}},
  \bibinfo {author} {\bibfnamefont {A.}~\bibnamefont {Keesling}}, \bibinfo
  {author} {\bibfnamefont {G.}~\bibnamefont {Semeghini}}, \bibinfo {author}
  {\bibfnamefont {T.~T.}\ \bibnamefont {Wang}}, \bibinfo {author}
  {\bibfnamefont {S.}~\bibnamefont {Ebadi}}, \bibinfo {author} {\bibfnamefont
  {H.}~\bibnamefont {Bernien}}, \bibinfo {author} {\bibfnamefont {A.~S.}\
  \bibnamefont {Zibrov}}, \bibinfo {author} {\bibfnamefont {H.}~\bibnamefont
  {Pichler}}, \bibinfo {author} {\bibfnamefont {S.}~\bibnamefont {Choi}},
  \bibinfo {author} {\bibfnamefont {J.}~\bibnamefont {Cui}}, \bibinfo {author}
  {\bibfnamefont {M.}~\bibnamefont {Rossignolo}}, \bibinfo {author}
  {\bibfnamefont {P.}~\bibnamefont {Rembold}}, \bibinfo {author} {\bibfnamefont
  {S.}~\bibnamefont {Montangero}}, \bibinfo {author} {\bibfnamefont
  {T.}~\bibnamefont {Calarco}}, \bibinfo {author} {\bibfnamefont
  {M.}~\bibnamefont {Endres}}, \bibinfo {author} {\bibfnamefont
  {M.}~\bibnamefont {Greiner}}, \bibinfo {author} {\bibfnamefont
  {V.}~\bibnamefont {Vuleti\'c}},\ and\ \bibinfo {author} {\bibfnamefont
  {M.~D.}\ \bibnamefont {Lukin}},\ }\bibfield  {title} {\bibinfo {title}
  {Generation and manipulation of Schr\"odinger cat states in {Rydberg} atom
  arrays},\ }\href {https://doi.org/10.1126/science.aax9743} {\bibfield
  {journal} {\bibinfo  {journal} {Science}\ }\textbf {\bibinfo {volume}
  {365}},\ \bibinfo {pages} {570} (\bibinfo {year} {2019})}\BibitemShut
  {NoStop}%
\bibitem [{\citenamefont {Song}\ \emph {et~al.}(2019)\citenamefont {Song},
  \citenamefont {Xu}, \citenamefont {Li}, \citenamefont {Zhang}, \citenamefont
  {Zhang}, \citenamefont {Liu}, \citenamefont {Guo}, \citenamefont {Wang},
  \citenamefont {Ren}, \citenamefont {Hao}, \citenamefont {Feng}, \citenamefont
  {Fan}, \citenamefont {Zheng}, \citenamefont {Wang}, \citenamefont {Wang},\
  and\ \citenamefont {Zhu}}]{Songetal2019}%
  \BibitemOpen
  \bibfield  {author} {\bibinfo {author} {\bibfnamefont {C.}~\bibnamefont
  {Song}}, \bibinfo {author} {\bibfnamefont {K.}~\bibnamefont {Xu}}, \bibinfo
  {author} {\bibfnamefont {H.}~\bibnamefont {Li}}, \bibinfo {author}
  {\bibfnamefont {Y.-R.}\ \bibnamefont {Zhang}}, \bibinfo {author}
  {\bibfnamefont {X.}~\bibnamefont {Zhang}}, \bibinfo {author} {\bibfnamefont
  {W.}~\bibnamefont {Liu}}, \bibinfo {author} {\bibfnamefont {Q.}~\bibnamefont
  {Guo}}, \bibinfo {author} {\bibfnamefont {Z.}~\bibnamefont {Wang}}, \bibinfo
  {author} {\bibfnamefont {W.}~\bibnamefont {Ren}}, \bibinfo {author}
  {\bibfnamefont {J.}~\bibnamefont {Hao}}, \bibinfo {author} {\bibfnamefont
  {H.}~\bibnamefont {Feng}}, \bibinfo {author} {\bibfnamefont {H.}~\bibnamefont
  {Fan}}, \bibinfo {author} {\bibfnamefont {D.}~\bibnamefont {Zheng}}, \bibinfo
  {author} {\bibfnamefont {D.-W.}\ \bibnamefont {Wang}}, \bibinfo {author}
  {\bibfnamefont {H.}~\bibnamefont {Wang}},\ and\ \bibinfo {author}
  {\bibfnamefont {S.-Y.}\ \bibnamefont {Zhu}},\ }\bibfield  {title} {\bibinfo
  {title} {Generation of multicomponent atomic Schr\"odinger cat states of up
  to 20 qubits},\ }\href {https://doi.org/10.1126/science.aay0600} {\bibfield
  {journal} {\bibinfo  {journal} {Science}\ }\textbf {\bibinfo {volume}
  {365}},\ \bibinfo {pages} {574} (\bibinfo {year} {2019})}\BibitemShut
  {NoStop}%
\bibitem [{\citenamefont {Mooney}\ \emph {et~al.}(2019)\citenamefont {Mooney},
  \citenamefont {Hill},\ and\ \citenamefont {Hollenberg}}]{Mooneyetal2019}%
  \BibitemOpen
  \bibfield  {author} {\bibinfo {author} {\bibfnamefont {G.~J.}\ \bibnamefont
  {Mooney}}, \bibinfo {author} {\bibfnamefont {C.~D.}\ \bibnamefont {Hill}},\
  and\ \bibinfo {author} {\bibfnamefont {L.~C.~L.}\ \bibnamefont
  {Hollenberg}},\ }\bibfield  {title} {\bibinfo {title} {Entanglement in a
  20-qubit superconducting quantum computer},\ }\href
  {https://doi.org/10.1038/s41598-019-49805-7} {\bibfield  {journal} {\bibinfo
  {journal} {Sci. Rep.}\ }\textbf {\bibinfo {volume} {9}},\ \bibinfo {pages}
  {13465} (\bibinfo {year} {2019})}\BibitemShut {NoStop}%
\bibitem [{\citenamefont {Keimer}\ and\ \citenamefont
  {Moore}(2017)}]{keimer2017physics}%
  \BibitemOpen
  \bibfield  {author} {\bibinfo {author} {\bibfnamefont {B.}~\bibnamefont
  {Keimer}}\ and\ \bibinfo {author} {\bibfnamefont {J.}~\bibnamefont {Moore}},\
  }\bibfield  {title} {\bibinfo {title} {The physics of quantum materials},\
  }\href {https://doi.org/10.1038/nphys4302} {\bibfield  {journal} {\bibinfo
  {journal} {Nat. Phys.}\ }\textbf {\bibinfo {volume} {13}},\ \bibinfo {pages}
  {1045} (\bibinfo {year} {2017})}\BibitemShut {NoStop}%
\bibitem [{\citenamefont {Tokura}\ \emph {et~al.}(2017)\citenamefont {Tokura},
  \citenamefont {Kawasaki},\ and\ \citenamefont
  {Nagaosa}}]{tokura2017emergent}%
  \BibitemOpen
  \bibfield  {author} {\bibinfo {author} {\bibfnamefont {Y.}~\bibnamefont
  {Tokura}}, \bibinfo {author} {\bibfnamefont {M.}~\bibnamefont {Kawasaki}},\
  and\ \bibinfo {author} {\bibfnamefont {N.}~\bibnamefont {Nagaosa}},\
  }\bibfield  {title} {\bibinfo {title} {Emergent functions of quantum
  materials},\ }\href {https://doi.org/10.1038/nphys4274} {\bibfield  {journal}
  {\bibinfo  {journal} {Nat. Phys.}\ }\textbf {\bibinfo {volume} {13}},\
  \bibinfo {pages} {1056} (\bibinfo {year} {2017})}\BibitemShut {NoStop}%
\bibitem [{\citenamefont {Adesso}\ \emph {et~al.}(2016)\citenamefont {Adesso},
  \citenamefont {Bromley},\ and\ \citenamefont {Cianciaruso}}]{AdessoBC2016}%
  \BibitemOpen
  \bibfield  {author} {\bibinfo {author} {\bibfnamefont {G.}~\bibnamefont
  {Adesso}}, \bibinfo {author} {\bibfnamefont {T.~R.}\ \bibnamefont
  {Bromley}},\ and\ \bibinfo {author} {\bibfnamefont {M.}~\bibnamefont
  {Cianciaruso}},\ }\bibfield  {title} {\bibinfo {title} {Measures and
  applications of quantum correlations},\ }\href
  {https://doi.org/10.1088/1751-8113/49/47/473001} {\bibfield  {journal}
  {\bibinfo  {journal} {J. Phys. A: Math. Theor.}\ }\textbf {\bibinfo {volume}
  {49}},\ \bibinfo {pages} {473001} (\bibinfo {year} {2016})}\BibitemShut
  {NoStop}%
\bibitem [{\citenamefont {Chiara}\ and\ \citenamefont
  {Sanpera}(2018)}]{DeChiara2018}%
  \BibitemOpen
  \bibfield  {author} {\bibinfo {author} {\bibfnamefont {G.~D.}\ \bibnamefont
  {Chiara}}\ and\ \bibinfo {author} {\bibfnamefont {A.}~\bibnamefont
  {Sanpera}},\ }\bibfield  {title} {\bibinfo {title} {Genuine quantum
  correlations in quantum many-body systems: a review of recent progress},\
  }\href {https://doi.org/10.1088/1361-6633/aabf61} {\bibfield  {journal}
  {\bibinfo  {journal} {Rep. Prog. Phys.}\ }\textbf {\bibinfo {volume} {81}},\
  \bibinfo {pages} {074002} (\bibinfo {year} {2018})}\BibitemShut {NoStop}%
\bibitem [{\citenamefont {Pezz\`e}\ \emph {et~al.}(2018)\citenamefont
  {Pezz\`e}, \citenamefont {Smerzi}, \citenamefont {Oberthaler}, \citenamefont
  {Schmied},\ and\ \citenamefont {Treutlein}}]{Pezzeetal2018}%
  \BibitemOpen
  \bibfield  {author} {\bibinfo {author} {\bibfnamefont {L.}~\bibnamefont
  {Pezz\`e}}, \bibinfo {author} {\bibfnamefont {A.}~\bibnamefont {Smerzi}},
  \bibinfo {author} {\bibfnamefont {M.~K.}\ \bibnamefont {Oberthaler}},
  \bibinfo {author} {\bibfnamefont {R.}~\bibnamefont {Schmied}},\ and\ \bibinfo
  {author} {\bibfnamefont {P.}~\bibnamefont {Treutlein}},\ }\bibfield  {title}
  {\bibinfo {title} {Quantum metrology with nonclassical states of atomic
  ensembles},\ }\href {https://doi.org/10.1103/RevModPhys.90.035005} {\bibfield
   {journal} {\bibinfo  {journal} {Rev. Mod. Phys.}\ }\textbf {\bibinfo
  {volume} {90}},\ \bibinfo {pages} {035005} (\bibinfo {year}
  {2018})}\BibitemShut {NoStop}%
\bibitem [{\citenamefont {Fr\'erot}\ \emph {et~al.}(2023)\citenamefont
  {Fr\'erot}, \citenamefont {Fadel},\ and\ \citenamefont
  {Lewenstein}}]{Frerot2023}%
  \BibitemOpen
  \bibfield  {author} {\bibinfo {author} {\bibfnamefont {I.}~\bibnamefont
  {Fr\'erot}}, \bibinfo {author} {\bibfnamefont {M.}~\bibnamefont {Fadel}},\
  and\ \bibinfo {author} {\bibfnamefont {M.}~\bibnamefont {Lewenstein}},\
  }\bibfield  {title} {\bibinfo {title} {Probing quantum correlations in
  many-body systems: a review of scalable methods},\ }\href
  {https://doi.org/10.1088/1361-6633/acf8d7} {\bibfield  {journal} {\bibinfo
  {journal} {Rep. Prog. Phys}\ }\textbf {\bibinfo {volume} {86}},\ \bibinfo
  {pages} {114001} (\bibinfo {year} {2023})}\BibitemShut {NoStop}%
\bibitem [{\citenamefont {Hauke}\ \emph {et~al.}(2016)\citenamefont {Hauke},
  \citenamefont {Heyl}, \citenamefont {Tagliacozzo},\ and\ \citenamefont
  {Zoller}}]{Hauke2016}%
  \BibitemOpen
  \bibfield  {author} {\bibinfo {author} {\bibfnamefont {P.}~\bibnamefont
  {Hauke}}, \bibinfo {author} {\bibfnamefont {M.}~\bibnamefont {Heyl}},
  \bibinfo {author} {\bibfnamefont {L.}~\bibnamefont {Tagliacozzo}},\ and\
  \bibinfo {author} {\bibfnamefont {P.}~\bibnamefont {Zoller}},\ }\bibfield
  {title} {\bibinfo {title} {Measuring multipartite entanglement through
  dynamic susceptibilities},\ }\href {https://doi.org/10.1038/nphys3700}
  {\bibfield  {journal} {\bibinfo  {journal} {Nat. Phys.}\ }\textbf {\bibinfo
  {volume} {12}},\ \bibinfo {pages} {778} (\bibinfo {year} {2016})}\BibitemShut
  {NoStop}%
\bibitem [{\citenamefont {Fr\'erot}\ and\ \citenamefont
  {Roscilde}(2016)}]{FrerotR2016}%
  \BibitemOpen
  \bibfield  {author} {\bibinfo {author} {\bibfnamefont {I.}~\bibnamefont
  {Fr\'erot}}\ and\ \bibinfo {author} {\bibfnamefont {T.}~\bibnamefont
  {Roscilde}},\ }\bibfield  {title} {\bibinfo {title} {Quantum variance: A
  measure of quantum coherence and quantum correlations for many-body
  systems},\ }\href {https://doi.org/10.1103/PhysRevB.94.075121} {\bibfield
  {journal} {\bibinfo  {journal} {Phys. Rev. B}\ }\textbf {\bibinfo {volume}
  {94}},\ \bibinfo {pages} {075121} (\bibinfo {year} {2016})}\BibitemShut
  {NoStop}%
\bibitem [{\citenamefont {Fr\'erot}\ and\ \citenamefont
  {Roscilde}(2019)}]{Frerot2019}%
  \BibitemOpen
  \bibfield  {author} {\bibinfo {author} {\bibfnamefont {I.}~\bibnamefont
  {Fr\'erot}}\ and\ \bibinfo {author} {\bibfnamefont {T.}~\bibnamefont
  {Roscilde}},\ }\bibfield  {title} {\bibinfo {title} {Reconstructing the
  quantum critical fan of strongly correlated systems using quantum
  correlations},\ }\href {https://doi.org/10.1038/s41467-019-08324-9}
  {\bibfield  {journal} {\bibinfo  {journal} {Nat. Commun.}\ }\textbf {\bibinfo
  {volume} {10}},\ \bibinfo {pages} {577} (\bibinfo {year} {2019})}\BibitemShut
  {NoStop}%
\bibitem [{\citenamefont {Mathew}\ \emph {et~al.}(2020)\citenamefont {Mathew},
  \citenamefont {Silva}, \citenamefont {Jain}, \citenamefont {Mohan},
  \citenamefont {Adroja}, \citenamefont {Sakai}, \citenamefont {Tomy},
  \citenamefont {Banerjee}, \citenamefont {Goreti}, \citenamefont {V.N.},
  \citenamefont {Singh},\ and\ \citenamefont
  {Jaiswal-Nagar}}]{PhysRevResearch.2.043329}%
  \BibitemOpen
  \bibfield  {author} {\bibinfo {author} {\bibfnamefont {G.}~\bibnamefont
  {Mathew}}, \bibinfo {author} {\bibfnamefont {S.~L.~L.}\ \bibnamefont
  {Silva}}, \bibinfo {author} {\bibfnamefont {A.}~\bibnamefont {Jain}},
  \bibinfo {author} {\bibfnamefont {A.}~\bibnamefont {Mohan}}, \bibinfo
  {author} {\bibfnamefont {D.~T.}\ \bibnamefont {Adroja}}, \bibinfo {author}
  {\bibfnamefont {V.~G.}\ \bibnamefont {Sakai}}, \bibinfo {author}
  {\bibfnamefont {C.~V.}\ \bibnamefont {Tomy}}, \bibinfo {author}
  {\bibfnamefont {A.}~\bibnamefont {Banerjee}}, \bibinfo {author}
  {\bibfnamefont {R.}~\bibnamefont {Goreti}}, \bibinfo {author} {\bibfnamefont
  {A.}~\bibnamefont {V.N.}}, \bibinfo {author} {\bibfnamefont {R.}~\bibnamefont
  {Singh}},\ and\ \bibinfo {author} {\bibfnamefont {D.}~\bibnamefont
  {Jaiswal-Nagar}},\ }\bibfield  {title} {\bibinfo {title} {Experimental
  realization of multipartite entanglement via quantum {Fisher} information in
  a uniform antiferromagnetic quantum spin chain},\ }\href
  {https://doi.org/10.1103/PhysRevResearch.2.043329} {\bibfield  {journal}
  {\bibinfo  {journal} {Phys. Rev. Research}\ }\textbf {\bibinfo {volume}
  {2}},\ \bibinfo {pages} {043329} (\bibinfo {year} {2020})}\BibitemShut
  {NoStop}%
\bibitem [{\citenamefont {Scheie}\ \emph {et~al.}(2021)\citenamefont {Scheie},
  \citenamefont {Laurell}, \citenamefont {Samarakoon}, \citenamefont {Lake},
  \citenamefont {Nagler}, \citenamefont {Granroth}, \citenamefont {Okamoto},
  \citenamefont {Alvarez},\ and\ \citenamefont
  {Tennant}}]{PhysRevB.103.224434}%
  \BibitemOpen
  \bibfield  {author} {\bibinfo {author} {\bibfnamefont {A.}~\bibnamefont
  {Scheie}}, \bibinfo {author} {\bibfnamefont {P.}~\bibnamefont {Laurell}},
  \bibinfo {author} {\bibfnamefont {A.~M.}\ \bibnamefont {Samarakoon}},
  \bibinfo {author} {\bibfnamefont {B.}~\bibnamefont {Lake}}, \bibinfo {author}
  {\bibfnamefont {S.~E.}\ \bibnamefont {Nagler}}, \bibinfo {author}
  {\bibfnamefont {G.~E.}\ \bibnamefont {Granroth}}, \bibinfo {author}
  {\bibfnamefont {S.}~\bibnamefont {Okamoto}}, \bibinfo {author} {\bibfnamefont
  {G.}~\bibnamefont {Alvarez}},\ and\ \bibinfo {author} {\bibfnamefont {D.~A.}\
  \bibnamefont {Tennant}},\ }\bibfield  {title} {\bibinfo {title} {Witnessing
  entanglement in quantum magnets using neutron scattering},\ }\href
  {https://doi.org/10.1103/PhysRevB.103.224434} {\bibfield  {journal} {\bibinfo
   {journal} {Phys. Rev. B}\ }\textbf {\bibinfo {volume} {103}},\ \bibinfo
  {pages} {224434} (\bibinfo {year} {2021})}\BibitemShut {NoStop}%
\bibitem [{\citenamefont {Laurell}\ \emph {et~al.}(2021)\citenamefont
  {Laurell}, \citenamefont {Scheie}, \citenamefont {Mukherjee}, \citenamefont
  {Koza}, \citenamefont {Enderle}, \citenamefont {Tylczynski}, \citenamefont
  {Okamoto}, \citenamefont {Coldea}, \citenamefont {Tennant},\ and\
  \citenamefont {Alvarez}}]{PhysRevLett.127.037201}%
  \BibitemOpen
  \bibfield  {author} {\bibinfo {author} {\bibfnamefont {P.}~\bibnamefont
  {Laurell}}, \bibinfo {author} {\bibfnamefont {A.}~\bibnamefont {Scheie}},
  \bibinfo {author} {\bibfnamefont {C.~J.}\ \bibnamefont {Mukherjee}}, \bibinfo
  {author} {\bibfnamefont {M.~M.}\ \bibnamefont {Koza}}, \bibinfo {author}
  {\bibfnamefont {M.}~\bibnamefont {Enderle}}, \bibinfo {author} {\bibfnamefont
  {Z.}~\bibnamefont {Tylczynski}}, \bibinfo {author} {\bibfnamefont
  {S.}~\bibnamefont {Okamoto}}, \bibinfo {author} {\bibfnamefont
  {R.}~\bibnamefont {Coldea}}, \bibinfo {author} {\bibfnamefont {D.~A.}\
  \bibnamefont {Tennant}},\ and\ \bibinfo {author} {\bibfnamefont
  {G.}~\bibnamefont {Alvarez}},\ }\bibfield  {title} {\bibinfo {title}
  {Quantifying and controlling entanglement in the quantum magnet
  {${\mathrm{Cs}}_{2}{\mathrm{CoCl}}_{4}$}},\ }\href
  {https://doi.org/10.1103/PhysRevLett.127.037201} {\bibfield  {journal}
  {\bibinfo  {journal} {Phys. Rev. Lett.}\ }\textbf {\bibinfo {volume} {127}},\
  \bibinfo {pages} {037201} (\bibinfo {year} {2021})}\BibitemShut {NoStop}%
\bibitem [{\citenamefont {Braunstein}\ and\ \citenamefont
  {Caves}(1994)}]{BraunsteinC1994}%
  \BibitemOpen
  \bibfield  {author} {\bibinfo {author} {\bibfnamefont {S.~L.}\ \bibnamefont
  {Braunstein}}\ and\ \bibinfo {author} {\bibfnamefont {C.~M.}\ \bibnamefont
  {Caves}},\ }\bibfield  {title} {\bibinfo {title} {Statistical distance and
  the geometry of quantum states},\ }\href
  {https://doi.org/10.1103/PhysRevLett.72.3439} {\bibfield  {journal} {\bibinfo
   {journal} {Phys. Rev. Lett.}\ }\textbf {\bibinfo {volume} {72}},\ \bibinfo
  {pages} {3439} (\bibinfo {year} {1994})}\BibitemShut {NoStop}%
\bibitem [{\citenamefont {Scheie}\ \emph {et~al.}(2024)\citenamefont {Scheie},
  \citenamefont {Ghioldi}, \citenamefont {Xing}, \citenamefont {Paddison},
  \citenamefont {Sherman}, \citenamefont {Dupont}, \citenamefont {Sanjeewa},
  \citenamefont {Lee}, \citenamefont {Woods}, \citenamefont {Abernathy},
  \citenamefont {Pajerowski}, \citenamefont {Williams}, \citenamefont {Zhang},
  \citenamefont {Manuel}, \citenamefont {Trumper}, \citenamefont {Pemmaraju},
  \citenamefont {Sefat}, \citenamefont {Parker}, \citenamefont {Devereaux},
  \citenamefont {Movshovich}, \citenamefont {Moore}, \citenamefont {Batista},\
  and\ \citenamefont {Tennant}}]{Scheie2021}%
  \BibitemOpen
  \bibfield  {author} {\bibinfo {author} {\bibfnamefont {A.~O.}\ \bibnamefont
  {Scheie}}, \bibinfo {author} {\bibfnamefont {E.~A.}\ \bibnamefont {Ghioldi}},
  \bibinfo {author} {\bibfnamefont {J.}~\bibnamefont {Xing}}, \bibinfo {author}
  {\bibfnamefont {J.~A.~M.}\ \bibnamefont {Paddison}}, \bibinfo {author}
  {\bibfnamefont {N.~E.}\ \bibnamefont {Sherman}}, \bibinfo {author}
  {\bibfnamefont {M.}~\bibnamefont {Dupont}}, \bibinfo {author} {\bibfnamefont
  {L.~D.}\ \bibnamefont {Sanjeewa}}, \bibinfo {author} {\bibfnamefont
  {S.}~\bibnamefont {Lee}}, \bibinfo {author} {\bibfnamefont {A.~J.}\
  \bibnamefont {Woods}}, \bibinfo {author} {\bibfnamefont {D.}~\bibnamefont
  {Abernathy}}, \bibinfo {author} {\bibfnamefont {D.~M.}\ \bibnamefont
  {Pajerowski}}, \bibinfo {author} {\bibfnamefont {T.~J.}\ \bibnamefont
  {Williams}}, \bibinfo {author} {\bibfnamefont {S.-S.}\ \bibnamefont {Zhang}},
  \bibinfo {author} {\bibfnamefont {L.~O.}\ \bibnamefont {Manuel}}, \bibinfo
  {author} {\bibfnamefont {A.~E.}\ \bibnamefont {Trumper}}, \bibinfo {author}
  {\bibfnamefont {C.~D.}\ \bibnamefont {Pemmaraju}}, \bibinfo {author}
  {\bibfnamefont {A.~S.}\ \bibnamefont {Sefat}}, \bibinfo {author}
  {\bibfnamefont {D.~S.}\ \bibnamefont {Parker}}, \bibinfo {author}
  {\bibfnamefont {T.~P.}\ \bibnamefont {Devereaux}}, \bibinfo {author}
  {\bibfnamefont {R.}~\bibnamefont {Movshovich}}, \bibinfo {author}
  {\bibfnamefont {J.~E.}\ \bibnamefont {Moore}}, \bibinfo {author}
  {\bibfnamefont {C.~D.}\ \bibnamefont {Batista}},\ and\ \bibinfo {author}
  {\bibfnamefont {D.~A.}\ \bibnamefont {Tennant}},\ }\bibfield  {title}
  {\bibinfo {title} {Proximate spin liquid and fractionalization in the
  triangular antiferromagnet {KYbSe$_2$}},\ }\href
  {https://doi.org/10.1038/s41567-023-02259-1} {\bibfield  {journal} {\bibinfo
  {journal} {Nature Physics}\ }\textbf {\bibinfo {volume} {20}},\ \bibinfo
  {pages} {74} (\bibinfo {year} {2024})}\BibitemShut {NoStop}%
\bibitem [{\citenamefont {Lake}\ \emph
  {et~al.}(2005{\natexlab{a}})\citenamefont {Lake}, \citenamefont {Tennant},
  \citenamefont {Frost},\ and\ \citenamefont {Nagler}}]{Lake2005}%
  \BibitemOpen
  \bibfield  {author} {\bibinfo {author} {\bibfnamefont {B.}~\bibnamefont
  {Lake}}, \bibinfo {author} {\bibfnamefont {D.~A.}\ \bibnamefont {Tennant}},
  \bibinfo {author} {\bibfnamefont {C.~D.}\ \bibnamefont {Frost}},\ and\
  \bibinfo {author} {\bibfnamefont {S.~E.}\ \bibnamefont {Nagler}},\ }\bibfield
   {title} {\bibinfo {title} {Quantum criticality and universal scaling of a
  quantum antiferromagnet},\ }\href {https://doi.org/10.1038/nmat1327}
  {\bibfield  {journal} {\bibinfo  {journal} {Nat. Mater.}\ }\textbf {\bibinfo
  {volume} {4}},\ \bibinfo {pages} {329} (\bibinfo {year}
  {2005}{\natexlab{a}})}\BibitemShut {NoStop}%
\bibitem [{\citenamefont {Lake}\ \emph {et~al.}(2013)\citenamefont {Lake},
  \citenamefont {Tennant}, \citenamefont {Caux}, \citenamefont {Barthel},
  \citenamefont {Schollw\"ock}, \citenamefont {Nagler},\ and\ \citenamefont
  {Frost}}]{PhysRevLett.111.137205}%
  \BibitemOpen
  \bibfield  {author} {\bibinfo {author} {\bibfnamefont {B.}~\bibnamefont
  {Lake}}, \bibinfo {author} {\bibfnamefont {D.~A.}\ \bibnamefont {Tennant}},
  \bibinfo {author} {\bibfnamefont {J.-S.}\ \bibnamefont {Caux}}, \bibinfo
  {author} {\bibfnamefont {T.}~\bibnamefont {Barthel}}, \bibinfo {author}
  {\bibfnamefont {U.}~\bibnamefont {Schollw\"ock}}, \bibinfo {author}
  {\bibfnamefont {S.~E.}\ \bibnamefont {Nagler}},\ and\ \bibinfo {author}
  {\bibfnamefont {C.~D.}\ \bibnamefont {Frost}},\ }\bibfield  {title} {\bibinfo
  {title} {Multispinon continua at zero and finite temperature in a near-ideal
  {Heisenberg} chain},\ }\href {https://doi.org/10.1103/PhysRevLett.111.137205}
  {\bibfield  {journal} {\bibinfo  {journal} {Phys. Rev. Lett.}\ }\textbf
  {\bibinfo {volume} {111}},\ \bibinfo {pages} {137205} (\bibinfo {year}
  {2013})}\BibitemShut {NoStop}%
\bibitem [{\citenamefont {Malpetti}\ and\ \citenamefont
  {Roscilde}(2016)}]{MalpettiR2016}%
  \BibitemOpen
  \bibfield  {author} {\bibinfo {author} {\bibfnamefont {D.}~\bibnamefont
  {Malpetti}}\ and\ \bibinfo {author} {\bibfnamefont {T.}~\bibnamefont
  {Roscilde}},\ }\bibfield  {title} {\bibinfo {title} {Quantum correlations,
  separability, and quantum coherence length in equilibrium many-body
  systems},\ }\href {https://doi.org/10.1103/PhysRevLett.117.130401} {\bibfield
   {journal} {\bibinfo  {journal} {Phys. Rev. Lett.}\ }\textbf {\bibinfo
  {volume} {117}},\ \bibinfo {pages} {130401} (\bibinfo {year}
  {2016})}\BibitemShut {NoStop}%
\bibitem [{\citenamefont {Kuwahara}\ and\ \citenamefont
  {Saito}(2022)}]{PhysRevX.12.021022}%
  \BibitemOpen
  \bibfield  {author} {\bibinfo {author} {\bibfnamefont {T.}~\bibnamefont
  {Kuwahara}}\ and\ \bibinfo {author} {\bibfnamefont {K.}~\bibnamefont
  {Saito}},\ }\bibfield  {title} {\bibinfo {title} {Exponential clustering of
  bipartite quantum entanglement at arbitrary temperatures},\ }\href
  {https://doi.org/10.1103/PhysRevX.12.021022} {\bibfield  {journal} {\bibinfo
  {journal} {Phys. Rev. X}\ }\textbf {\bibinfo {volume} {12}},\ \bibinfo
  {pages} {021022} (\bibinfo {year} {2022})}\BibitemShut {NoStop}%
\bibitem [{\citenamefont {Wigner}\ and\ \citenamefont
  {Yanase}(1963)}]{Wigner_1963}%
  \BibitemOpen
  \bibfield  {author} {\bibinfo {author} {\bibfnamefont {E.~P.}\ \bibnamefont
  {Wigner}}\ and\ \bibinfo {author} {\bibfnamefont {M.~M.}\ \bibnamefont
  {Yanase}},\ }\bibfield  {title} {\bibinfo {title} {Information contents of
  distributions},\ }\href {https://doi.org/10.1073/pnas.49.6.910} {\bibfield
  {journal} {\bibinfo  {journal} {Proc. Natl. Acad. Sci. U.S.A.}\ }\textbf
  {\bibinfo {volume} {49}},\ \bibinfo {pages} {910} (\bibinfo {year}
  {1963})}\BibitemShut {NoStop}%
\bibitem [{\citenamefont {Forster}(1995)}]{forster_book}%
  \BibitemOpen
  \bibfield  {author} {\bibinfo {author} {\bibfnamefont {D.}~\bibnamefont
  {Forster}},\ }\href@noop {} {\emph {\bibinfo {title} {Hydrodynamic
  Fluctuations, Broken Symmetry, and Correlation Functions}}},\ Advanced book
  classics\ (\bibinfo  {publisher} {Avalon Publishing},\ \bibinfo {year}
  {1995})\BibitemShut {NoStop}%
\bibitem [{\citenamefont {Petz}(1996)}]{Petz1996}%
  \BibitemOpen
  \bibfield  {author} {\bibinfo {author} {\bibfnamefont {D.}~\bibnamefont
  {Petz}},\ }\bibfield  {title} {\bibinfo {title} {Monotone metrics on matrix
  spaces},\ }\href
  {https://doi.org/http://dx.doi.org/10.1016/0024-3795(94)00211-8} {\bibfield
  {journal} {\bibinfo  {journal} {Linear Algebra Appl.}\ }\textbf {\bibinfo
  {volume} {244}},\ \bibinfo {pages} {81 } (\bibinfo {year}
  {1996})}\BibitemShut {NoStop}%
\bibitem [{\citenamefont {Gibilisco}\ \emph {et~al.}(2009)\citenamefont
  {Gibilisco}, \citenamefont {Imparato},\ and\ \citenamefont
  {Isola}}]{Gibiliscoetal2007}%
  \BibitemOpen
  \bibfield  {author} {\bibinfo {author} {\bibfnamefont {P.}~\bibnamefont
  {Gibilisco}}, \bibinfo {author} {\bibfnamefont {D.}~\bibnamefont
  {Imparato}},\ and\ \bibinfo {author} {\bibfnamefont {T.}~\bibnamefont
  {Isola}},\ }\bibfield  {title} {\bibinfo {title} {Inequalities for quantum
  fisher information},\ }\href
  {https://doi.org/https://doi.org/10.1090/S0002-9939-08-09447-1} {\bibfield
  {journal} {\bibinfo  {journal} {Proc. Amer. Math. Soc.}\ }\textbf {\bibinfo
  {volume} {137}},\ \bibinfo {pages} {317} (\bibinfo {year}
  {2009})}\BibitemShut {NoStop}%
\bibitem [{\citenamefont {Fr\'erot}(2017)}]{Frerot2017}%
  \BibitemOpen
  \bibfield  {author} {\bibinfo {author} {\bibfnamefont {I.}~\bibnamefont
  {Fr\'erot}},\ }\emph {\bibinfo {title} {A quantum statistical approach to
  quantum correlations in many-body systems}},\ \href
  {https://tel.archives-ouvertes.fr/tel-01679743} {\bibinfo {type} {{Ph.D.}
  thesis}},\ \bibinfo  {school} {\'Ecole normale sup\'erieure de Lyon}
  (\bibinfo {year} {2017})\BibitemShut {NoStop}%
\bibitem [{\citenamefont {Fr\'erot}\ \emph
  {et~al.}(2022{\natexlab{a}})\citenamefont {Fr\'erot}, \citenamefont
  {Ran\ifmmode~\mbox{\c{c}}\else \c{c}\fi{}on},\ and\ \citenamefont
  {Roscilde}}]{Frerotetal2022}%
  \BibitemOpen
  \bibfield  {author} {\bibinfo {author} {\bibfnamefont {I.}~\bibnamefont
  {Fr\'erot}}, \bibinfo {author} {\bibfnamefont {A.}~\bibnamefont
  {Ran\ifmmode~\mbox{\c{c}}\else \c{c}\fi{}on}},\ and\ \bibinfo {author}
  {\bibfnamefont {T.}~\bibnamefont {Roscilde}},\ }\bibfield  {title} {\bibinfo
  {title} {Thermal critical dynamics from equilibrium quantum fluctuations},\
  }\href {https://doi.org/10.1103/PhysRevLett.128.130601} {\bibfield  {journal}
  {\bibinfo  {journal} {Phys. Rev. Lett.}\ }\textbf {\bibinfo {volume} {128}},\
  \bibinfo {pages} {130601} (\bibinfo {year} {2022}{\natexlab{a}})}\BibitemShut
  {NoStop}%
\bibitem [{\citenamefont {Liu}\ \emph {et~al.}(2019)\citenamefont {Liu},
  \citenamefont {Yuan}, \citenamefont {Lu},\ and\ \citenamefont
  {Wang}}]{Liuetal2020}%
  \BibitemOpen
  \bibfield  {author} {\bibinfo {author} {\bibfnamefont {J.}~\bibnamefont
  {Liu}}, \bibinfo {author} {\bibfnamefont {H.}~\bibnamefont {Yuan}}, \bibinfo
  {author} {\bibfnamefont {X.-M.}\ \bibnamefont {Lu}},\ and\ \bibinfo {author}
  {\bibfnamefont {X.}~\bibnamefont {Wang}},\ }\bibfield  {title} {\bibinfo
  {title} {Quantum Fisher information matrix and multiparameter estimation},\
  }\href {https://doi.org/10.1088/1751-8121/ab5d4d} {\bibfield  {journal}
  {\bibinfo  {journal} {J. Phys. A: Math. Theor.}\ }\textbf {\bibinfo {volume}
  {53}},\ \bibinfo {pages} {023001} (\bibinfo {year} {2019})}\BibitemShut
  {NoStop}%
\bibitem [{\citenamefont {Jarrell}\ and\ \citenamefont
  {Gubernatis}(1996)}]{JARRELL1996}%
  \BibitemOpen
  \bibfield  {author} {\bibinfo {author} {\bibfnamefont {M.}~\bibnamefont
  {Jarrell}}\ and\ \bibinfo {author} {\bibfnamefont {J.}~\bibnamefont
  {Gubernatis}},\ }\bibfield  {title} {\bibinfo {title} {Bayesian inference and
  the analytic continuation of imaginary-time quantum Monte Carlo data},\
  }\href {https://doi.org/https://doi.org/10.1016/0370-1573(95)00074-7}
  {\bibfield  {journal} {\bibinfo  {journal} {Phys. Rep.}\ }\textbf {\bibinfo
  {volume} {269}},\ \bibinfo {pages} {133} (\bibinfo {year}
  {1996})}\BibitemShut {NoStop}%
\bibitem [{\citenamefont {Lovesey}(1984)}]{Lovesey1984}%
  \BibitemOpen
  \bibfield  {author} {\bibinfo {author} {\bibfnamefont {S.}~\bibnamefont
  {Lovesey}},\ }\href@noop {} {\emph {\bibinfo {title} {Theory of Neutron
  Scattering from Condensed Matter}}}\ (\bibinfo  {publisher} {Clarendon
  Press},\ \bibinfo {address} {Oxford},\ \bibinfo {year} {1984})\BibitemShut
  {NoStop}%
\bibitem [{\citenamefont {Scheie}\ \emph {et~al.}(2022)\citenamefont {Scheie},
  \citenamefont {Laurell}, \citenamefont {Lake}, \citenamefont {Nagler},
  \citenamefont {Stone}, \citenamefont {Caux},\ and\ \citenamefont
  {Tennant}}]{Scheie2022}%
  \BibitemOpen
  \bibfield  {author} {\bibinfo {author} {\bibfnamefont {A.}~\bibnamefont
  {Scheie}}, \bibinfo {author} {\bibfnamefont {P.}~\bibnamefont {Laurell}},
  \bibinfo {author} {\bibfnamefont {B.}~\bibnamefont {Lake}}, \bibinfo {author}
  {\bibfnamefont {S.~E.}\ \bibnamefont {Nagler}}, \bibinfo {author}
  {\bibfnamefont {M.~B.}\ \bibnamefont {Stone}}, \bibinfo {author}
  {\bibfnamefont {J.-S.}\ \bibnamefont {Caux}},\ and\ \bibinfo {author}
  {\bibfnamefont {D.~A.}\ \bibnamefont {Tennant}},\ }\bibfield  {title}
  {\bibinfo {title} {Quantum wake dynamics in {Heisenberg} antiferromagnetic
  chains},\ }\href {https://doi.org/10.1038/s41467-022-33571-8} {\bibfield
  {journal} {\bibinfo  {journal} {Nat. Commun.}\ }\textbf {\bibinfo {volume}
  {13}},\ \bibinfo {pages} {5796} (\bibinfo {year} {2022})}\BibitemShut
  {NoStop}%
\bibitem [{\citenamefont {Tennant}\ \emph {et~al.}(1993)\citenamefont
  {Tennant}, \citenamefont {Perring}, \citenamefont {Cowley},\ and\
  \citenamefont {Nagler}}]{PhysRevLett.70.4003}%
  \BibitemOpen
  \bibfield  {author} {\bibinfo {author} {\bibfnamefont {D.~A.}\ \bibnamefont
  {Tennant}}, \bibinfo {author} {\bibfnamefont {T.~G.}\ \bibnamefont
  {Perring}}, \bibinfo {author} {\bibfnamefont {R.~A.}\ \bibnamefont
  {Cowley}},\ and\ \bibinfo {author} {\bibfnamefont {S.~E.}\ \bibnamefont
  {Nagler}},\ }\bibfield  {title} {\bibinfo {title} {Unbound spinons in the
  {S=1/2} antiferromagnetic chain {${\mathrm{KCuF}}_{3}$}},\ }\href
  {https://doi.org/10.1103/PhysRevLett.70.4003} {\bibfield  {journal} {\bibinfo
   {journal} {Phys. Rev. Lett.}\ }\textbf {\bibinfo {volume} {70}},\ \bibinfo
  {pages} {4003} (\bibinfo {year} {1993})}\BibitemShut {NoStop}%
\bibitem [{\citenamefont {Lake}\ \emph
  {et~al.}(2005{\natexlab{b}})\citenamefont {Lake}, \citenamefont {Tennant},\
  and\ \citenamefont {Nagler}}]{PhysRevB.71.134412}%
  \BibitemOpen
  \bibfield  {author} {\bibinfo {author} {\bibfnamefont {B.}~\bibnamefont
  {Lake}}, \bibinfo {author} {\bibfnamefont {D.~A.}\ \bibnamefont {Tennant}},\
  and\ \bibinfo {author} {\bibfnamefont {S.~E.}\ \bibnamefont {Nagler}},\
  }\bibfield  {title} {\bibinfo {title} {Longitudinal magnetic dynamics and
  dimensional crossover in the quasi-one-dimensional spin-$\frac{1}{2}$
  {Heisenberg} antiferromagnet {${\mathrm{KCuF}}_{3}$}},\ }\href
  {https://doi.org/10.1103/PhysRevB.71.134412} {\bibfield  {journal} {\bibinfo
  {journal} {Phys. Rev. B}\ }\textbf {\bibinfo {volume} {71}},\ \bibinfo
  {pages} {134412} (\bibinfo {year} {2005}{\natexlab{b}})}\BibitemShut
  {NoStop}%
\bibitem [{\citenamefont {Squires}(2012)}]{Squires}%
  \BibitemOpen
  \bibfield  {author} {\bibinfo {author} {\bibfnamefont {G.~L.}\ \bibnamefont
  {Squires}},\ }\href@noop {} {\emph {\bibinfo {title} {Introduction to the
  Theory of Thermal Neutron Scattering}}},\ \bibinfo {edition} {3rd}\ ed.\
  (\bibinfo  {publisher} {Cambridge University Press},\ \bibinfo {address}
  {Cambridge, UK},\ \bibinfo {year} {2012})\BibitemShut {NoStop}%
\bibitem [{\citenamefont {Sylju\aa{}sen}\ and\ \citenamefont
  {Sandvik}(2002)}]{SyljuasenS2002}%
  \BibitemOpen
  \bibfield  {author} {\bibinfo {author} {\bibfnamefont {O.~F.}\ \bibnamefont
  {Sylju\aa{}sen}}\ and\ \bibinfo {author} {\bibfnamefont {A.~W.}\ \bibnamefont
  {Sandvik}},\ }\bibfield  {title} {\bibinfo {title} {Quantum Monte Carlo with
  directed loops},\ }\href {https://doi.org/10.1103/PhysRevE.66.046701}
  {\bibfield  {journal} {\bibinfo  {journal} {Phys. Rev. E}\ }\textbf {\bibinfo
  {volume} {66}},\ \bibinfo {pages} {046701} (\bibinfo {year}
  {2002})}\BibitemShut {NoStop}%
\bibitem [{\citenamefont {Hyllus}\ \emph {et~al.}(2012)\citenamefont {Hyllus},
  \citenamefont {Laskowski}, \citenamefont {Krischek}, \citenamefont
  {Schwemmer}, \citenamefont {Wieczorek}, \citenamefont {Weinfurter},
  \citenamefont {Pezz\'e},\ and\ \citenamefont {Smerzi}}]{Hyllus2012}%
  \BibitemOpen
  \bibfield  {author} {\bibinfo {author} {\bibfnamefont {P.}~\bibnamefont
  {Hyllus}}, \bibinfo {author} {\bibfnamefont {W.}~\bibnamefont {Laskowski}},
  \bibinfo {author} {\bibfnamefont {R.}~\bibnamefont {Krischek}}, \bibinfo
  {author} {\bibfnamefont {C.}~\bibnamefont {Schwemmer}}, \bibinfo {author}
  {\bibfnamefont {W.}~\bibnamefont {Wieczorek}}, \bibinfo {author}
  {\bibfnamefont {H.}~\bibnamefont {Weinfurter}}, \bibinfo {author}
  {\bibfnamefont {L.}~\bibnamefont {Pezz\'e}},\ and\ \bibinfo {author}
  {\bibfnamefont {A.}~\bibnamefont {Smerzi}},\ }\bibfield  {title} {\bibinfo
  {title} {Fisher information and multiparticle entanglement},\ }\href
  {https://doi.org/10.1103/PhysRevA.85.022321} {\bibfield  {journal} {\bibinfo
  {journal} {Phys. Rev. A}\ }\textbf {\bibinfo {volume} {85}},\ \bibinfo
  {pages} {022321} (\bibinfo {year} {2012})}\BibitemShut {NoStop}%
\bibitem [{\citenamefont {T\'oth}(2012)}]{Toth2012}%
  \BibitemOpen
  \bibfield  {author} {\bibinfo {author} {\bibfnamefont {G.}~\bibnamefont
  {T\'oth}},\ }\bibfield  {title} {\bibinfo {title} {Multipartite entanglement
  and high-precision metrology},\ }\href
  {https://doi.org/10.1103/PhysRevA.85.022322} {\bibfield  {journal} {\bibinfo
  {journal} {Phys. Rev. A}\ }\textbf {\bibinfo {volume} {85}},\ \bibinfo
  {pages} {022322} (\bibinfo {year} {2012})}\BibitemShut {NoStop}%
\bibitem [{\citenamefont {Yasuda}\ \emph {et~al.}(2005)\citenamefont {Yasuda},
  \citenamefont {Todo}, \citenamefont {Hukushima}, \citenamefont {Alet},
  \citenamefont {Keller}, \citenamefont {Troyer},\ and\ \citenamefont
  {Takayama}}]{Yasudaetal2005}%
  \BibitemOpen
  \bibfield  {author} {\bibinfo {author} {\bibfnamefont {C.}~\bibnamefont
  {Yasuda}}, \bibinfo {author} {\bibfnamefont {S.}~\bibnamefont {Todo}},
  \bibinfo {author} {\bibfnamefont {K.}~\bibnamefont {Hukushima}}, \bibinfo
  {author} {\bibfnamefont {F.}~\bibnamefont {Alet}}, \bibinfo {author}
  {\bibfnamefont {M.}~\bibnamefont {Keller}}, \bibinfo {author} {\bibfnamefont
  {M.}~\bibnamefont {Troyer}},\ and\ \bibinfo {author} {\bibfnamefont
  {H.}~\bibnamefont {Takayama}},\ }\bibfield  {title} {\bibinfo {title} {N\'eel
  temperature of quasi-low-dimensional Heisenberg antiferromagnets},\ }\href
  {https://doi.org/10.1103/PhysRevLett.94.217201} {\bibfield  {journal}
  {\bibinfo  {journal} {Phys. Rev. Lett.}\ }\textbf {\bibinfo {volume} {94}},\
  \bibinfo {pages} {217201} (\bibinfo {year} {2005})}\BibitemShut {NoStop}%
\bibitem [{\citenamefont {Coffman}\ \emph {et~al.}(2000)\citenamefont
  {Coffman}, \citenamefont {Kundu},\ and\ \citenamefont
  {Wootters}}]{PhysRevA.61.052306}%
  \BibitemOpen
  \bibfield  {author} {\bibinfo {author} {\bibfnamefont {V.}~\bibnamefont
  {Coffman}}, \bibinfo {author} {\bibfnamefont {J.}~\bibnamefont {Kundu}},\
  and\ \bibinfo {author} {\bibfnamefont {W.~K.}\ \bibnamefont {Wootters}},\
  }\bibfield  {title} {\bibinfo {title} {Distributed entanglement},\ }\href
  {https://doi.org/10.1103/PhysRevA.61.052306} {\bibfield  {journal} {\bibinfo
  {journal} {Phys. Rev. A}\ }\textbf {\bibinfo {volume} {61}},\ \bibinfo
  {pages} {052306} (\bibinfo {year} {2000})}\BibitemShut {NoStop}%
\bibitem [{\citenamefont {Osborne}\ and\ \citenamefont
  {Verstraete}(2006)}]{PhysRevLett.96.220503}%
  \BibitemOpen
  \bibfield  {author} {\bibinfo {author} {\bibfnamefont {T.~J.}\ \bibnamefont
  {Osborne}}\ and\ \bibinfo {author} {\bibfnamefont {F.}~\bibnamefont
  {Verstraete}},\ }\bibfield  {title} {\bibinfo {title} {General monogamy
  inequality for bipartite qubit entanglement},\ }\href
  {https://doi.org/10.1103/PhysRevLett.96.220503} {\bibfield  {journal}
  {\bibinfo  {journal} {Phys. Rev. Lett.}\ }\textbf {\bibinfo {volume} {96}},\
  \bibinfo {pages} {220503} (\bibinfo {year} {2006})}\BibitemShut {NoStop}%
\bibitem [{\citenamefont {Sch{\"u}lke}(2007)}]{schulke2007electron}%
  \BibitemOpen
  \bibfield  {author} {\bibinfo {author} {\bibfnamefont {W.}~\bibnamefont
  {Sch{\"u}lke}},\ }\href@noop {} {\emph {\bibinfo {title} {Electron Dynamics
  by Inelastic X-ray Scattering}}},\ \bibinfo {series} {Oxford Series on
  Synchrotron Radiation}, Vol.~\bibinfo {volume} {7}\ (\bibinfo  {publisher}
  {Oxford University Press},\ \bibinfo {year} {2007})\BibitemShut {NoStop}%
\bibitem [{\citenamefont {Vig}\ \emph {et~al.}(2017)\citenamefont {Vig},
  \citenamefont {Kogar}, \citenamefont {Mitrano}, \citenamefont {Husain},
  \citenamefont {Mishra}, \citenamefont {Rak}, \citenamefont {Venema},
  \citenamefont {Johnson}, \citenamefont {Gu}, \citenamefont {Fradkin},
  \citenamefont {Norman},\ and\ \citenamefont
  {Abbamonte}}]{SciPostPhys.3.4.026}%
  \BibitemOpen
  \bibfield  {author} {\bibinfo {author} {\bibfnamefont {S.}~\bibnamefont
  {Vig}}, \bibinfo {author} {\bibfnamefont {A.}~\bibnamefont {Kogar}}, \bibinfo
  {author} {\bibfnamefont {M.}~\bibnamefont {Mitrano}}, \bibinfo {author}
  {\bibfnamefont {A.~A.}\ \bibnamefont {Husain}}, \bibinfo {author}
  {\bibfnamefont {V.}~\bibnamefont {Mishra}}, \bibinfo {author} {\bibfnamefont
  {M.~S.}\ \bibnamefont {Rak}}, \bibinfo {author} {\bibfnamefont
  {L.}~\bibnamefont {Venema}}, \bibinfo {author} {\bibfnamefont {P.~D.}\
  \bibnamefont {Johnson}}, \bibinfo {author} {\bibfnamefont {G.~D.}\
  \bibnamefont {Gu}}, \bibinfo {author} {\bibfnamefont {E.}~\bibnamefont
  {Fradkin}}, \bibinfo {author} {\bibfnamefont {M.~R.}\ \bibnamefont
  {Norman}},\ and\ \bibinfo {author} {\bibfnamefont {P.}~\bibnamefont
  {Abbamonte}},\ }\bibfield  {title} {\bibinfo {title} {{Measurement of the
  dynamic charge response of materials using low-energy, momentum-resolved
  electron energy-loss spectroscopy (M-EELS)}},\ }\href
  {https://doi.org/10.21468/SciPostPhys.3.4.026} {\bibfield  {journal}
  {\bibinfo  {journal} {SciPost Phys.}\ }\textbf {\bibinfo {volume} {3}},\
  \bibinfo {pages} {026} (\bibinfo {year} {2017})}\BibitemShut {NoStop}%
\bibitem [{\citenamefont {Gegenwart}\ \emph {et~al.}(2008)\citenamefont
  {Gegenwart}, \citenamefont {Si},\ and\ \citenamefont
  {Steglich}}]{gegenwart2008quantum}%
  \BibitemOpen
  \bibfield  {author} {\bibinfo {author} {\bibfnamefont {P.}~\bibnamefont
  {Gegenwart}}, \bibinfo {author} {\bibfnamefont {Q.}~\bibnamefont {Si}},\ and\
  \bibinfo {author} {\bibfnamefont {F.}~\bibnamefont {Steglich}},\ }\bibfield
  {title} {\bibinfo {title} {Quantum criticality in heavy-fermion metals},\
  }\href {https://doi.org/10.1038/nphys892} {\bibfield  {journal} {\bibinfo
  {journal} {Nat. Phys.}\ }\textbf {\bibinfo {volume} {4}},\ \bibinfo {pages}
  {186} (\bibinfo {year} {2008})}\BibitemShut {NoStop}%
\bibitem [{\citenamefont {Gabbrielli}\ \emph {et~al.}(2018)\citenamefont
  {Gabbrielli}, \citenamefont {Smerzi},\ and\ \citenamefont
  {{Pezz{\`e}}}}]{Gabbriellietal2018}%
  \BibitemOpen
  \bibfield  {author} {\bibinfo {author} {\bibfnamefont {M.}~\bibnamefont
  {Gabbrielli}}, \bibinfo {author} {\bibfnamefont {A.}~\bibnamefont {Smerzi}},\
  and\ \bibinfo {author} {\bibfnamefont {L.}~\bibnamefont {{Pezz{\`e}}}},\
  }\bibfield  {title} {\bibinfo {title} {Multipartite {Entanglement} at
  {Finite} {Temperature}},\ }\href {https://doi.org/10.1038/s41598-018-31761-3}
  {\bibfield  {journal} {\bibinfo  {journal} {Sci. Rep.}\ }\textbf {\bibinfo
  {volume} {8}},\ \bibinfo {pages} {15663} (\bibinfo {year}
  {2018})}\BibitemShut {NoStop}%
\bibitem [{\citenamefont {Coldea}\ \emph {et~al.}(2002)\citenamefont {Coldea},
  \citenamefont {Tennant}, \citenamefont {Habicht}, \citenamefont {Smeibidl},
  \citenamefont {Wolters},\ and\ \citenamefont
  {Tylczynski}}]{PhysRevLett.88.137203}%
  \BibitemOpen
  \bibfield  {author} {\bibinfo {author} {\bibfnamefont {R.}~\bibnamefont
  {Coldea}}, \bibinfo {author} {\bibfnamefont {D.~A.}\ \bibnamefont {Tennant}},
  \bibinfo {author} {\bibfnamefont {K.}~\bibnamefont {Habicht}}, \bibinfo
  {author} {\bibfnamefont {P.}~\bibnamefont {Smeibidl}}, \bibinfo {author}
  {\bibfnamefont {C.}~\bibnamefont {Wolters}},\ and\ \bibinfo {author}
  {\bibfnamefont {Z.}~\bibnamefont {Tylczynski}},\ }\bibfield  {title}
  {\bibinfo {title} {Direct measurement of the spin hamiltonian and observation
  of condensation of magnons in the 2d frustrated quantum magnet
  {${\mathrm{Cs}}_{2}{\mathrm{CuCl}}_{4}$}},\ }\href
  {https://doi.org/10.1103/PhysRevLett.88.137203} {\bibfield  {journal}
  {\bibinfo  {journal} {Phys. Rev. Lett.}\ }\textbf {\bibinfo {volume} {88}},\
  \bibinfo {pages} {137203} (\bibinfo {year} {2002})}\BibitemShut {NoStop}%
\bibitem [{\citenamefont {Starykh}\ \emph {et~al.}(2010)\citenamefont
  {Starykh}, \citenamefont {Katsura},\ and\ \citenamefont
  {Balents}}]{PhysRevB.82.014421}%
  \BibitemOpen
  \bibfield  {author} {\bibinfo {author} {\bibfnamefont {O.~A.}\ \bibnamefont
  {Starykh}}, \bibinfo {author} {\bibfnamefont {H.}~\bibnamefont {Katsura}},\
  and\ \bibinfo {author} {\bibfnamefont {L.}~\bibnamefont {Balents}},\
  }\bibfield  {title} {\bibinfo {title} {Extreme sensitivity of a frustrated
  quantum magnet: {${\mathrm{Cs}}_{2}\mathrm{Cu}{\mathrm{Cl}}_{4}$}},\ }\href
  {https://doi.org/10.1103/PhysRevB.82.014421} {\bibfield  {journal} {\bibinfo
  {journal} {Phys. Rev. B}\ }\textbf {\bibinfo {volume} {82}},\ \bibinfo
  {pages} {014421} (\bibinfo {year} {2010})}\BibitemShut {NoStop}%
\bibitem [{\citenamefont {McKay}\ and\ \citenamefont
  {DeMarco}(2011)}]{McKay2011}%
  \BibitemOpen
  \bibfield  {author} {\bibinfo {author} {\bibfnamefont {D.~C.}\ \bibnamefont
  {McKay}}\ and\ \bibinfo {author} {\bibfnamefont {B.}~\bibnamefont
  {DeMarco}},\ }\bibfield  {title} {\bibinfo {title} {Cooling in strongly
  correlated optical lattices: prospects and challenges},\ }\href
  {https://doi.org/10.1088/0034-4885/74/5/054401} {\bibfield  {journal}
  {\bibinfo  {journal} {Rep. Prog. Phys.}\ }\textbf {\bibinfo {volume} {74}},\
  \bibinfo {pages} {054401} (\bibinfo {year} {2011})}\BibitemShut {NoStop}%
\bibitem [{\citenamefont {Trotzky}\ \emph {et~al.}(2010)\citenamefont
  {Trotzky}, \citenamefont {Pollet}, \citenamefont {Gerbier}, \citenamefont
  {Schnorrberger}, \citenamefont {Bloch}, \citenamefont {Prokof'ev},
  \citenamefont {Svistunov},\ and\ \citenamefont {Troyer}}]{Trotzky2010}%
  \BibitemOpen
  \bibfield  {author} {\bibinfo {author} {\bibfnamefont {S.}~\bibnamefont
  {Trotzky}}, \bibinfo {author} {\bibfnamefont {L.}~\bibnamefont {Pollet}},
  \bibinfo {author} {\bibfnamefont {F.}~\bibnamefont {Gerbier}}, \bibinfo
  {author} {\bibfnamefont {U.}~\bibnamefont {Schnorrberger}}, \bibinfo {author}
  {\bibfnamefont {I.}~\bibnamefont {Bloch}}, \bibinfo {author} {\bibfnamefont
  {N.~V.}\ \bibnamefont {Prokof'ev}}, \bibinfo {author} {\bibfnamefont
  {B.}~\bibnamefont {Svistunov}},\ and\ \bibinfo {author} {\bibfnamefont
  {M.}~\bibnamefont {Troyer}},\ }\bibfield  {title} {\bibinfo {title}
  {Suppression of the critical temperature for superfluidity near the Mott
  transition},\ }\href {https://doi.org/10.1038/nphys1799} {\bibfield
  {journal} {\bibinfo  {journal} {Nat. Phys.}\ }\textbf {\bibinfo {volume}
  {6}},\ \bibinfo {pages} {998} (\bibinfo {year} {2010})}\BibitemShut {NoStop}%
\bibitem [{\citenamefont {Carcy}\ \emph {et~al.}(2021)\citenamefont {Carcy},
  \citenamefont {Herc\'e}, \citenamefont {Tenart}, \citenamefont {Roscilde},\
  and\ \citenamefont {Cl\'ement}}]{Carcy2021}%
  \BibitemOpen
  \bibfield  {author} {\bibinfo {author} {\bibfnamefont {C.}~\bibnamefont
  {Carcy}}, \bibinfo {author} {\bibfnamefont {G.}~\bibnamefont {Herc\'e}},
  \bibinfo {author} {\bibfnamefont {A.}~\bibnamefont {Tenart}}, \bibinfo
  {author} {\bibfnamefont {T.}~\bibnamefont {Roscilde}},\ and\ \bibinfo
  {author} {\bibfnamefont {D.}~\bibnamefont {Cl\'ement}},\ }\bibfield  {title}
  {\bibinfo {title} {Certifying the adiabatic preparation of ultracold lattice
  bosons in the vicinity of the Mott transition},\ }\href
  {https://doi.org/10.1103/PhysRevLett.126.045301} {\bibfield  {journal}
  {\bibinfo  {journal} {Phys. Rev. Lett.}\ }\textbf {\bibinfo {volume} {126}},\
  \bibinfo {pages} {045301} (\bibinfo {year} {2021})}\BibitemShut {NoStop}%
\bibitem [{\citenamefont {Egelstaff}\ and\ \citenamefont
  {Pease}(1954)}]{egelstaff1954design}%
  \BibitemOpen
  \bibfield  {author} {\bibinfo {author} {\bibfnamefont {P.}~\bibnamefont
  {Egelstaff}}\ and\ \bibinfo {author} {\bibfnamefont {R.}~\bibnamefont
  {Pease}},\ }\bibfield  {title} {\bibinfo {title} {The design of cold neutron
  filters},\ }\href {https://doi.org/10.1088/0950-7671/31/6/305} {\bibfield
  {journal} {\bibinfo  {journal} {J. Sci. Instrum.}\ }\textbf {\bibinfo
  {volume} {31}},\ \bibinfo {pages} {207} (\bibinfo {year} {1954})}\BibitemShut
  {NoStop}%
\bibitem [{\citenamefont {Tennant}(1988)}]{tennant1988performance}%
  \BibitemOpen
  \bibfield  {author} {\bibinfo {author} {\bibfnamefont {D.}~\bibnamefont
  {Tennant}},\ }\bibfield  {title} {\bibinfo {title} {Performance of a cooled
  sapphire and beryllium assembly for filtering of thermal neutrons},\ }\href
  {https://doi.org/10.1063/1.1140212} {\bibfield  {journal} {\bibinfo
  {journal} {Rev. Sci. Instrum.}\ }\textbf {\bibinfo {volume} {59}},\ \bibinfo
  {pages} {380} (\bibinfo {year} {1988})}\BibitemShut {NoStop}%
\bibitem [{\citenamefont {Mourigal}\ \emph {et~al.}(2013)\citenamefont
  {Mourigal}, \citenamefont {Enderle}, \citenamefont {Kl\"opperpieper},
  \citenamefont {Caux}, \citenamefont {Stunault},\ and\ \citenamefont
  {R\o{}nnow}}]{Mourigal2013}%
  \BibitemOpen
  \bibfield  {author} {\bibinfo {author} {\bibfnamefont {M.}~\bibnamefont
  {Mourigal}}, \bibinfo {author} {\bibfnamefont {M.}~\bibnamefont {Enderle}},
  \bibinfo {author} {\bibfnamefont {A.}~\bibnamefont {Kl\"opperpieper}},
  \bibinfo {author} {\bibfnamefont {J.-S.}\ \bibnamefont {Caux}}, \bibinfo
  {author} {\bibfnamefont {A.}~\bibnamefont {Stunault}},\ and\ \bibinfo
  {author} {\bibfnamefont {H.~M.}\ \bibnamefont {R\o{}nnow}},\ }\bibfield
  {title} {\bibinfo {title} {Fractional spinon excitations in the quantum
  {Heisenberg} antiferromagnetic chain},\ }\href
  {https://doi.org/10.1038/nphys2652} {\bibfield  {journal} {\bibinfo
  {journal} {Nat. Phys.}\ }\textbf {\bibinfo {volume} {9}},\ \bibinfo {pages}
  {435} (\bibinfo {year} {2013})}\BibitemShut {NoStop}%
\bibitem [{\citenamefont {Wu}\ \emph {et~al.}(2019)\citenamefont {Wu},
  \citenamefont {Nikitin}, \citenamefont {Wang}, \citenamefont {Zhu},
  \citenamefont {Batista}, \citenamefont {Tsvelik}, \citenamefont {Samarakoon},
  \citenamefont {Tennant}, \citenamefont {Brando}, \citenamefont {Vasylechko},
  \citenamefont {Frontzek}, \citenamefont {Savici}, \citenamefont {Sala},
  \citenamefont {Ehlers}, \citenamefont {Christianson}, \citenamefont
  {Lumsden},\ and\ \citenamefont {Podlesnyak}}]{Wu2019}%
  \BibitemOpen
  \bibfield  {author} {\bibinfo {author} {\bibfnamefont {L.~S.}\ \bibnamefont
  {Wu}}, \bibinfo {author} {\bibfnamefont {S.~E.}\ \bibnamefont {Nikitin}},
  \bibinfo {author} {\bibfnamefont {Z.}~\bibnamefont {Wang}}, \bibinfo {author}
  {\bibfnamefont {W.}~\bibnamefont {Zhu}}, \bibinfo {author} {\bibfnamefont
  {C.~D.}\ \bibnamefont {Batista}}, \bibinfo {author} {\bibfnamefont {A.~M.}\
  \bibnamefont {Tsvelik}}, \bibinfo {author} {\bibfnamefont {A.~M.}\
  \bibnamefont {Samarakoon}}, \bibinfo {author} {\bibfnamefont {D.~A.}\
  \bibnamefont {Tennant}}, \bibinfo {author} {\bibfnamefont {M.}~\bibnamefont
  {Brando}}, \bibinfo {author} {\bibfnamefont {L.}~\bibnamefont {Vasylechko}},
  \bibinfo {author} {\bibfnamefont {M.}~\bibnamefont {Frontzek}}, \bibinfo
  {author} {\bibfnamefont {A.~T.}\ \bibnamefont {Savici}}, \bibinfo {author}
  {\bibfnamefont {G.}~\bibnamefont {Sala}}, \bibinfo {author} {\bibfnamefont
  {G.}~\bibnamefont {Ehlers}}, \bibinfo {author} {\bibfnamefont {A.~D.}\
  \bibnamefont {Christianson}}, \bibinfo {author} {\bibfnamefont {M.~D.}\
  \bibnamefont {Lumsden}},\ and\ \bibinfo {author} {\bibfnamefont
  {A.}~\bibnamefont {Podlesnyak}},\ }\bibfield  {title} {\bibinfo {title}
  {{Tomonaga-Luttinger} liquid behavior and spinon confinement in
  {YbAlO}$_3$},\ }\href {https://doi.org/10.1038/s41467-019-08485-7} {\bibfield
   {journal} {\bibinfo  {journal} {Nat. Commun.}\ }\textbf {\bibinfo {volume}
  {10}},\ \bibinfo {pages} {698} (\bibinfo {year} {2019})}\BibitemShut
  {NoStop}%
\bibitem [{\citenamefont {Xu}\ \emph {et~al.}(2013)\citenamefont {Xu},
  \citenamefont {Xu},\ and\ \citenamefont {Tranquada}}]{Xu_AbsUnits_2013}%
  \BibitemOpen
  \bibfield  {author} {\bibinfo {author} {\bibfnamefont {G.}~\bibnamefont
  {Xu}}, \bibinfo {author} {\bibfnamefont {Z.}~\bibnamefont {Xu}},\ and\
  \bibinfo {author} {\bibfnamefont {J.~M.}\ \bibnamefont {Tranquada}},\
  }\bibfield  {title} {\bibinfo {title} {{Absolute cross-section normalization
  of magnetic neutron scattering data}},\ }\href
  {https://doi.org/10.1063/1.4818323} {\bibfield  {journal} {\bibinfo
  {journal} {Review of Scientific Instruments}\ }\textbf {\bibinfo {volume}
  {84}},\ \bibinfo {pages} {083906} (\bibinfo {year} {2013})}\BibitemShut
  {NoStop}%
\bibitem [{\citenamefont {Granroth}\ \emph {et~al.}(2010)\citenamefont
  {Granroth}, \citenamefont {Kolesnikov}, \citenamefont {Sherline},
  \citenamefont {Clancy}, \citenamefont {Ross}, \citenamefont {Ruff},
  \citenamefont {Gaulin},\ and\ \citenamefont {Nagler}}]{Granroth_2010}%
  \BibitemOpen
  \bibfield  {author} {\bibinfo {author} {\bibfnamefont {G.~E.}\ \bibnamefont
  {Granroth}}, \bibinfo {author} {\bibfnamefont {A.~I.}\ \bibnamefont
  {Kolesnikov}}, \bibinfo {author} {\bibfnamefont {T.~E.}\ \bibnamefont
  {Sherline}}, \bibinfo {author} {\bibfnamefont {J.~P.}\ \bibnamefont
  {Clancy}}, \bibinfo {author} {\bibfnamefont {K.~A.}\ \bibnamefont {Ross}},
  \bibinfo {author} {\bibfnamefont {J.~P.~C.}\ \bibnamefont {Ruff}}, \bibinfo
  {author} {\bibfnamefont {B.~D.}\ \bibnamefont {Gaulin}},\ and\ \bibinfo
  {author} {\bibfnamefont {S.~E.}\ \bibnamefont {Nagler}},\ }\bibfield  {title}
  {\bibinfo {title} {{SEQUOIA}: A newly operating chopper spectrometer at the
  {SNS}},\ }\href {https://doi.org/10.1088/1742-6596/251/1/012058} {\bibfield
  {journal} {\bibinfo  {journal} {J. Phys.: Conf. Ser.}\ }\textbf {\bibinfo
  {volume} {251}},\ \bibinfo {pages} {012058} (\bibinfo {year}
  {2010})}\BibitemShut {NoStop}%
\bibitem [{\citenamefont {Perring}\ \emph {et~al.}(1994)\citenamefont
  {Perring}, \citenamefont {Taylor}, \citenamefont {Osborn}, \citenamefont
  {Paul}, \citenamefont {Boothroyd},\ and\ \citenamefont
  {Aeppli}}]{perring1994proceedings}%
  \BibitemOpen
  \bibfield  {author} {\bibinfo {author} {\bibfnamefont {T.}~\bibnamefont
  {Perring}}, \bibinfo {author} {\bibfnamefont {A.}~\bibnamefont {Taylor}},
  \bibinfo {author} {\bibfnamefont {R.}~\bibnamefont {Osborn}}, \bibinfo
  {author} {\bibfnamefont {D.}~\bibnamefont {Paul}}, \bibinfo {author}
  {\bibfnamefont {A.}~\bibnamefont {Boothroyd}},\ and\ \bibinfo {author}
  {\bibfnamefont {G.}~\bibnamefont {Aeppli}},\ }\bibfield  {title} {\bibinfo
  {title} {in {Proceedings} of {ICANS XII}},\ }\href@noop {} {\bibfield
  {journal} {\bibinfo  {journal} {RAL Report}\ }\textbf {\bibinfo {volume}
  {94}},\ \bibinfo {pages} {1} (\bibinfo {year} {1994})}\BibitemShut {NoStop}%
\bibitem [{\citenamefont {Ikeda}\ and\ \citenamefont
  {Hirakawa}(1973)}]{ikeda1973neutron}%
  \BibitemOpen
  \bibfield  {author} {\bibinfo {author} {\bibfnamefont {H.}~\bibnamefont
  {Ikeda}}\ and\ \bibinfo {author} {\bibfnamefont {K.}~\bibnamefont
  {Hirakawa}},\ }\bibfield  {title} {\bibinfo {title} {Neutron diffraction
  study in one-dimensional antiferromagnet {KCuF}$_3$},\ }\href
  {https://doi.org/10.1143/JPSJ.35.722} {\bibfield  {journal} {\bibinfo
  {journal} {J. Phys. Soc. Jpn.}\ }\textbf {\bibinfo {volume} {35}},\ \bibinfo
  {pages} {722} (\bibinfo {year} {1973})}\BibitemShut {NoStop}%
\bibitem [{\citenamefont {White}(1992)}]{PhysRevLett.69.2863}%
  \BibitemOpen
  \bibfield  {author} {\bibinfo {author} {\bibfnamefont {S.~R.}\ \bibnamefont
  {White}},\ }\bibfield  {title} {\bibinfo {title} {Density matrix formulation
  for quantum renormalization groups},\ }\href
  {https://doi.org/10.1103/PhysRevLett.69.2863} {\bibfield  {journal} {\bibinfo
   {journal} {Phys. Rev. Lett.}\ }\textbf {\bibinfo {volume} {69}},\ \bibinfo
  {pages} {2863} (\bibinfo {year} {1992})}\BibitemShut {NoStop}%
\bibitem [{\citenamefont {White}(1993)}]{PhysRevB.48.10345}%
  \BibitemOpen
  \bibfield  {author} {\bibinfo {author} {\bibfnamefont {S.~R.}\ \bibnamefont
  {White}},\ }\bibfield  {title} {\bibinfo {title} {Density-matrix algorithms
  for quantum renormalization groups},\ }\href
  {https://doi.org/10.1103/PhysRevB.48.10345} {\bibfield  {journal} {\bibinfo
  {journal} {Phys. Rev. B}\ }\textbf {\bibinfo {volume} {48}},\ \bibinfo
  {pages} {10345} (\bibinfo {year} {1993})}\BibitemShut {NoStop}%
\bibitem [{\citenamefont {Feiguin}\ and\ \citenamefont
  {White}(2005)}]{PhysRevB.72.220401}%
  \BibitemOpen
  \bibfield  {author} {\bibinfo {author} {\bibfnamefont {A.~E.}\ \bibnamefont
  {Feiguin}}\ and\ \bibinfo {author} {\bibfnamefont {S.~R.}\ \bibnamefont
  {White}},\ }\bibfield  {title} {\bibinfo {title} {Finite-temperature density
  matrix renormalization using an enlarged Hilbert space},\ }\href
  {https://doi.org/10.1103/PhysRevB.72.220401} {\bibfield  {journal} {\bibinfo
  {journal} {Phys. Rev. B}\ }\textbf {\bibinfo {volume} {72}},\ \bibinfo
  {pages} {220401(R)} (\bibinfo {year} {2005})}\BibitemShut {NoStop}%
\bibitem [{\citenamefont {Barthel}\ \emph {et~al.}(2009)\citenamefont
  {Barthel}, \citenamefont {Schollw\"ock},\ and\ \citenamefont
  {White}}]{PhysRevB.79.245101}%
  \BibitemOpen
  \bibfield  {author} {\bibinfo {author} {\bibfnamefont {T.}~\bibnamefont
  {Barthel}}, \bibinfo {author} {\bibfnamefont {U.}~\bibnamefont
  {Schollw\"ock}},\ and\ \bibinfo {author} {\bibfnamefont {S.~R.}\ \bibnamefont
  {White}},\ }\bibfield  {title} {\bibinfo {title} {Spectral functions in
  one-dimensional quantum systems at finite temperature using the density
  matrix renormalization group},\ }\href
  {https://doi.org/10.1103/PhysRevB.79.245101} {\bibfield  {journal} {\bibinfo
  {journal} {Phys. Rev. B}\ }\textbf {\bibinfo {volume} {79}},\ \bibinfo
  {pages} {245101} (\bibinfo {year} {2009})}\BibitemShut {NoStop}%
\bibitem [{\citenamefont {Alvarez}(2009)}]{Alvarez2009}%
  \BibitemOpen
  \bibfield  {author} {\bibinfo {author} {\bibfnamefont {G.}~\bibnamefont
  {Alvarez}},\ }\bibfield  {title} {\bibinfo {title} {The density matrix
  renormalization group for strongly correlated electron systems: A generic
  implementation},\ }\href {https://doi.org/10.1016/j.cpc.2009.02.016}
  {\bibfield  {journal} {\bibinfo  {journal} {Comp. Phys. Comms.}\ }\textbf
  {\bibinfo {volume} {180}},\ \bibinfo {pages} {1572} (\bibinfo {year}
  {2009})}\BibitemShut {NoStop}%
\bibitem [{\citenamefont {K\"uhner}\ and\ \citenamefont
  {White}(1999)}]{PhysRevB.60.335}%
  \BibitemOpen
  \bibfield  {author} {\bibinfo {author} {\bibfnamefont {T.~D.}\ \bibnamefont
  {K\"uhner}}\ and\ \bibinfo {author} {\bibfnamefont {S.~R.}\ \bibnamefont
  {White}},\ }\bibfield  {title} {\bibinfo {title} {Dynamical correlation
  functions using the density matrix renormalization group},\ }\href
  {https://doi.org/10.1103/PhysRevB.60.335} {\bibfield  {journal} {\bibinfo
  {journal} {Phys. Rev. B}\ }\textbf {\bibinfo {volume} {60}},\ \bibinfo
  {pages} {335} (\bibinfo {year} {1999})}\BibitemShut {NoStop}%
\bibitem [{\citenamefont {Nocera}\ and\ \citenamefont
  {Alvarez}(2016)}]{PhysRevE.94.053308}%
  \BibitemOpen
  \bibfield  {author} {\bibinfo {author} {\bibfnamefont {A.}~\bibnamefont
  {Nocera}}\ and\ \bibinfo {author} {\bibfnamefont {G.}~\bibnamefont
  {Alvarez}},\ }\bibfield  {title} {\bibinfo {title} {Spectral functions with
  the density matrix renormalization group: Krylov-space approach for
  correction vectors},\ }\href {https://doi.org/10.1103/PhysRevE.94.053308}
  {\bibfield  {journal} {\bibinfo  {journal} {Phys. Rev. E}\ }\textbf {\bibinfo
  {volume} {94}},\ \bibinfo {pages} {053308} (\bibinfo {year}
  {2016})}\BibitemShut {NoStop}%
\end{thebibliography}

%

\end{document}